# Temperature and field dependence of the phase separation, structure, and magnetic ordering in $La_{1-x}Ca_xMnO_3$, (x=0.47, 0.50, and 0.53)

Q. Huang[1, 2], J. W. Lynn[1, 3], R. W. Erwin[1], A. Santoro[1], D. C. Dender[1,2], V. N. Smolyaninova[3], K. Ghosh[3], and R. L. Greene[3]

[1]*NIST Center for Neutron Research, National Institute of Standards and Technology, Gaithersburg, MD 20899*
[2]*Department of Materials and Nuclear Engineering, University of Maryland, College Park, MD 20742*
[3]*Center for Superconductivity Research, Department of Physics, University of Maryland, College Park, MD 20742*

## Abstract

Neutron powder diffraction measurements, combined with magnetization and resistivity data, have been carried out in the doped perovskite $La_{1-x}Ca_xMnO_3$ (x=0.47, 0.50, and 0.53) to elucidate the structural, magnetic, and electronic properties of the system around the composition corresponding to an equal number of $Mn^{3+}$ and $Mn^{4+}$. At room temperature all three samples are paramagnetic and single phase, with crystallographic symmetry *Pnma*. The samples then all become ferromagnetic (FM) at $T_C \approx 265$ K. At ~230 K, however, a second distinct crystallographic phase (denoted A-II) begins to form. Initially the intrinsic widths of the peaks are quite large, but they narrow as the temperature decreases and the phase fraction increases, indicating microscopic coexistence. The A-II phase has the same space-group symmetry and similar lattice parameters as the initial (F-I) phase. Both phases begin to exhibit weak Jahn-Teller distortions of the $MnO_6$ octahedra below ~230K, but the distortions are different in the two phases and vary in character with temperature, and can be interpreted in terms of ordering of the $\mathbf{d}_z^2$ orbitals. The fraction of the sample that exhibits the A-II phase increases with decreasing temperature and also increases with increasing Ca doping, but the transition never goes to completion to the lowest temperatures measured (5 K) and the two phases therefore coexist in this temperature-composition regime. Phase A-II orders antiferromagnetically (AFM) below a Néel temperature $T_N \approx 160$ K, with the CE-type magnetic structure. For the *x*=0.47 sample the proportion of F-I at 10 K is about 60%, which has afforded us the opportunity to fully characterize the crystallographic and magnetic behavior of both phases over the entire temperature range. The F-I phase maintains the ferromagnetic state for all temperatures below $T_C$, with the moment direction remaining along the *c*-axis. Resistivity measurements show that this phase is a conductor, while the CE phase is insulating. Application of magnetic fields up to 9T progressively inhibits the formation of the A-II phase, but this suppression is path dependent, being much stronger for example if the sample is field-cooled compared to zero-field cooling and then applying the field. The H-T phase diagram obtained from the diffraction measurements is in good agreement with the results of magnetization and resistivity. Overall the measurements underscore the delicate energetic balance between the magnetic, structural and electronic properties of the system.

## Introduction

The mixed-valent manganites $Ln_{1-x}B_xMnO_3$ (Ln=rare earth, B=Ca, Sr, or Ba) exhibit rich and interesting physical properties because of the strong interplay between lattice distortions, transport properties, and magnetic ordering.[1-8] The ground state of the $La_{1-x}Ca_xMnO_3$ system of particular interest here is either a ferromagnetic (FM) metal or an antiferromagnetic (AF) insulator depending on the composition. The $x$=0.5 system straddles the boundary separating these two ground states and has consequently been the focal point for investigations using a variety of experimental techniques, including x-ray and neutron powder diffraction techniques. Radaelli and co-workers[7,8] found that $La_{½}Ca_{½}MnO_3$ is a paramagnetic insulator at high temperatures, becoming ferromagnetic at ~225 K and then antiferromagnetic at ~155 K. The unusual broadening of the diffraction peaks between these temperatures was attributed by the authors to a distribution of lattice parameters in their sample, and they were in fact able to fit the high-resolution synchrotron x-ray patterns assuming the existence of four similar phases with slightly different lattice constants. The temperature dependence of the lattice parameters shows an increase of the *a* and *c* axes and a sharp decrease of *b* as the temperature decreases from 225 to 150 K. This behavior was interpreted as an indication of orbital ordering, with the $d_z^2$ orbital of $Mn^{3+}$ lying in the *a-c* plane of the structure. The magnetic ordering below 155 K was found to be of the CE-type, implying also the presence of charge ordering according to the structure proposed by Goodenough.[9] Electron microscopy studies by Mori et al.[10] showed that between 135 and 95 K, $La_{½}Ca_{½}MnO_3$ is a mixture of microdomains of (i) an AF, charge-ordered, incommensurate phase; (ii) a FM, charge disordered phase. The sizes of the microdomains are a function of temperature, being about 100-200 Å at 200 K and 500-600 Å at 125 K. Below 95 K, this mixture changes into a homogeneous AF commensurate phase with charge and orbital ordering. The existence of ferro- and antiferromagnetic domains associated with metallic and charge-ordered insulating phases, respectively, was also found by Allodi et al.[11] in an NMR study of the $x$=0.5 composition. Such domains were found to coexist at all temperatures below the magnetic transition, although their relative proportions were found to vary with temperature.

In all of these studies it was recognized that the FM and AFM modifications coexist in varying proportions at low temperatures, but it did not prove feasible to fully characterize the electronic, magnetic and structural behavior of these phases. In order to clarify these aspects of the $La_{1-x}Ca_xMnO_3$ system, we have carried out neutron powder diffraction, resistivity, and magnetization experiments for three compositions over a wider composition range near and including $x$=0.5.

## Experimental Procedures

Three polycrystalline samples of $La_{1-x}Ca_xMnO_3$ with $x$=0.47, 0.50, and 0.53 were prepared from stoichiometric amounts of $La_2O_3$, $CaCO_3$, and $MnCO_3$ by a standard solid state reaction technique.[12] The Ca and $Mn^{4+}$ concentrations were checked with wavelength-dispersive x-ray absorption (WDX) and titration experiments, which gave the nominal x values within ±0.01. Magnetization measurements were made with a commercial SQUID magnetometer, and the resistivity was measured by a standard four-probe technique.

The neutron-powder-diffraction intensity data were collected using the BT-1 high-resolution powder diffractometer located at NIST Center for Neutron Research. Cu(311) and Ge(311) monochromators were employed to produce monochromatic neutron beams of wavelengths 1.5401 Å and 2.0784 Å, respectively. The longer wavelength of 2.0784 Å produced by Ge(311) was used to obtain a higher neutron intensity and to achieve better resolution in the low 2θ angular range, which is particularly useful for investigating the magnetic structure and other temperature dependent properties. Collimators with horizontal divergences of 15′, 20′, and 7′ full width at half maximum (FWHM) were used before and after the monochromator, and after the sample, respectively. The intensities were measured in steps of 0.05° in the 2θ range 3°-168°. Data were collected at various temperatures in the range 10-300 K to determine the nuclear and magnetic structures and to detect and elucidate possible magnetic and structural transitions. The nuclear and magnetic parameters were refined by using the GSAS program.[13] The neutron scattering amplitudes used in the refinements were 0.827, 0.490, -0.373, and 0.581 ($\times 10^{-12}$ cm) for La, Ca, Mn, and O, respectively.[13] Models used in the initial stages of the refinements were those reported in Ref. 8. Field-dependent measurements were made by in a 9T vertical-field superconducting magnet and using only the Ge(311) as monochromator.

## Results

### *Crystal and Magnetic Structures*

The three samples all undergo the same sequence of structural and magnetic transitions as a function of temperature, and the overall phase diagram determined in the present study is presented in Fig. 1. At room temperature we find that all three samples are paramagnetic and exhibit the orthorhombic structure with *Pnma* symmetry. The lattice parameters are related to those of the basic perovskite unit cell as indicated in Fig. 1. Typical refinement results based on intensity data collected with both the Cu(311) and Ge(311) monochromators are given (for x=0.47) in Tables 1 and 2 at 295 and 10 K. The statistical uncertainties quoted represent one standard deviation.

A portion of the diffraction pattern for the x=0.47 sample is shown in Fig. 2 in the 2 θ region between 15° and 40°. The top part is for a temperature of 295 K, where we find that the sample is single-phase with Bragg peaks that are resolution-limited, as is the case for all three compositions. No impurity phases of any kind were detectable, attesting to the high quality of the samples. Upon cooling to 175 K we find that a ferromagnetic ordering has developed, and we will refer to the magnetic/structural phase in which the ferromagnetism occurs as the F-I phase in the following discussion. Structurally this is the phase observed at room temperature. Since the magnetic and nuclear unit cells are the same for ferromagnetic ordering and the magnetic peaks are then superposed on the nuclear peaks, the F-I ferromagnetic peaks are best revealed in the difference plot of Fig. 2b, obtained by subtracting from the observed intensities the profile calculated without the FM contribution.[14] The magnetic moment is along the *c*-axis, and we find no evidence of any kind of spin canting over the full temperature and composition range where ferromagnetism occurs.

Fig. 2c shows the data obtained for this sample at 10 K. The solid curve represents the contribution of the crystal structure, and the difference plot locates the magnetic reflections. We see that now we have two types of magnetic reflections, the ferromagnetic ones observed as in Fig. 2b, and a new set of antiferromagnetic peaks that belong to a distinct crystallographic phase. This new crystallographic phase begins to develop at ~230 K as discussed in detail below, while the antiferromagnetic order develops at ~160 K. We refer to this new phase as A-II. The fraction of F-I to A-II phases is temperature dependent, but the transformation from F-I to A-II is never complete down to a temperature of 5 K. The two phases therefore coexist, as is clearly evident in Fig. 2d where we find that a good fit of the data is obtained by taking into account both the FM and AFM contributions.

An example of a full diffraction pattern and the agreement between the observed and calculated intensities is shown in Fig. 3, which is for the $x$=0.50 composition at 10 K. This sample becomes ferromagnetic below 265 K. At ~230 K the A-II structural phase begins to form, and the development of this new structural phase is shown more clearly in the inset. At room temperature a single peak is observed, which is a combination of the unresolved (031), (211) and (112) nuclear structure peaks for the F-I phase. For both the F-I and A-II phases there is no magnetic contribution in this angular region, so this scattering provides a good indication of the development and relative fraction of the two phases. The inset shows that at 250 K there is little if any A-II phase, but at 200 K there is clearly a second peak, which progressively develops and separates with further decrease of temperature. This high angle peak is the (031) Bragg reflection for the A-II phase, while the (211) and (112) peaks for the A-II phase are contained in the lower angle peak, along with all three peaks for the F-I phase. A careful inspection of the data at 175 K in Fig. 2b also reveals additional structural scattering, which is evident for example on the high-angle side of the peak at ~38 °.

In the temperature interval between 230 and 160 K the neutron diffraction peaks are considerably broadened, in agreement with the results of Radaelli and co-workers mentioned previously. Fig. 4a shows the width (FWHM) of the nuclear reflection $(031)_{A\text{-}II}$ and Fig. 4b its integrated intensity as a function of temperature. We see that the A-II phase develops in three stages upon cooling. From 230 to ~200 K, the proportion of A-II increases from zero to 14, 40, and 70% for $x$=0.47, 0.50, and 0.53, respectively. In this interval the FWHM of the $(031)_{A\text{-}II}$ remains practically unchanged at ~0.55°, which is about three times the width of the instrumental resolution. This intrinsic peak width of 0.52° then indicates a domain size of ~180 Å. From 200 to 160 K the FWHM decreases from 0.55° to 0.32° indicating that the domain size has doubled to ~360 Å, while the relative proportion of A-II increases only slightly for all three samples. From 160 K to 100 K the proportion of A-II increases again to the maximum values of 42%, 78%, and 90% for $x$=0.47, 0.50, and 0.53, respectively, while the width remains essentially constant.

The development of the A-II phase as observed in the neutron data corresponds nicely to the behavior found in resistivity and bulk magnetization measurements, as shown in Fig. 5. The low temperature (1-25 K) specific heat of these samples has been measured and reported elsewhere.[12] The resistivity is much lower for the x=0.47 sample than for the higher doping levels, clearly indicating that the ferromagnetic phase is metallic while the antiferromagnetic phase exhibits insulating behavior. There is also a direct correspondence between the fraction of the ferromagnetic phase and the bulk magnetization values obtained at low temperature. It is evident from the drop in the magnetization at lower temperatures, however, that the F-I phase is transforming into the A-II phase as the temperature is decreased. This transformation is clearly discontinuous, as indicated by the thermal hysteresis in the resistivity, magnetization, and phase fraction for the three samples.

The structure refinements for the F-I show that the magnetic moments are aligned along the *c* axis of the unit cell for all three compositions. This ordering and this orientation remain the same down to 5 K. This result is particularly evident in the case of the sample with $x$=0.47, where, at low temperatures, the proportion of F-I and A-II phases is about the same (see Fig. 2c and 2d). The temperature dependence of the ordered magnetic moments obtained from the refinements are shown in Fig. 6 for all three samples, where we find that the behavior is rather typical for magnetic order parameters, with ferromagnetic Curie points of 260, 265, and 240 K for $x$=0.47, 0.50, and 0.53, respectively, while the antiferromagnetic Néel temperature is 160 K.

The antiferromagnetic structure in the A-II phase is of the CE-type, with a unit cell (see Fig 1) four times larger than that of the nuclear structure. The symmetry and the initial orientation of the moments for the refinements of this structure were taken from the data reported in Table V of ref. 8. In this configuration the $Mn^{3+}$ and $Mn^{4+}$ cations are ordered and are located on two crystallographically independent sites and, as also noted in ref. 8, the magnetic reflections associated with the $Mn^{3+}$ cations are considerably broader than those generated by the $Mn^{4+}$ sublattice. This behavior has been convincingly explained there by assuming the existence of twinning generated by an operation that leaves the $Mn^{4+}$ magnetic sublattice unperturbed while it decreases the size of the $Mn^{3+}$ magnetic domains. As shown in Table 1, the orientation of the $Mn^{3+}$ magnetic moments is tilted towards the *a*-axis direction, consistent with the influence of the $\mathbf{d}_\mathbf{z}^2$ orbitals to align the moment along the [101] direction (as discussed below), while that of the $Mn^{4+}$ moments is parallel to the direction of the *c*-axis.

Since the $Mn^{3+}$ cations have four unpaired electrons and the $Mn^{4+}$ have only three, we should expect a larger magnetic moment for $Mn^{3+}$ than for $Mn^{4+}$. As shown in Fig. 6, however, this is not what we observe. For $x$=0.47 and 0.5 the magnitude of the $Mn^{3+}$ and $Mn^{4+}$ magnetic moments is practically the same at all temperatures, while for $x$=0.53 the $Mn^{3+}$ moment becomes smaller than that of $Mn^{4+}$ as the temperature decreases. These results may be explained as due to the smaller size of the $Mn^{3+}$ domains. Fits of the 110 magnetic reflection, generated by the $Mn^{3+}$ sublattice, show that its FWHM increases as the Ca doping increases, while the size of the $Mn^{4+}$ domains remains practically unaltered. The $Mn^{3+}$ domains therefore decrease in size as $x$ increases, with a consequent increase in the relative fraction of surface spins and, presumably, also an increase in the spin disorder in these domains increases, thus lowering the apparent $Mn^{3+}$ magnetic moment. The magnetic moments in both the F-I and A-II phases decrease as the Ca-doping increases, which is consistent with the theoretical predictions of Goodenough.[9]

### ***Crystal structure distortions***

#### *A-II Phase*

Plots of the lattice parameters and Mn-O distances as function of the temperature are shown in Figs. 7 and 8, respectively, for the $x$=0.50 composition; data for the $x$=0.47 and 0.53 samples show similar behavior and for clarity have been omitted from the figure. As the temperature decreases from ~200 K to 150 K, the *a* and *c* parameters increase and the *b* parameter decreases. This result, first reported by Radaelli and co-workers,[8] indicates that orbital ordering takes place in the *a-c* plane of the structure. The behavior of the Jahn-Teller distortion of the $MnO_6$ octahedra as a function of the temperature is more complex. As shown in Fig 9a (and in Figs. 8a and c), at 175 K (i.e. before the appearance of antiferromagnetic order and in the temperature range in which the diffraction peaks are broadened), the $MnO_6$ octahedra in the A-II phase are stretched in the $[10\bar{1}]$ direction and compressed along [010] and [101], indicating that the $d_z^2$ orbitals are all more or less oriented along $[10\bar{1}]$. According to Goodenough's model of the magnetic coupling in this class of materials, this orbital configuration would result in a C-type magnetic structure below the Néel temperature. Since this magnetic and orbital ordering would generate a configuration with high elastic energy [9], a re-orientation of the $d_z^2$ orbitals, consistent with the CE-type structure, is expected to take place at a temperature near the Néel temperature. In this new configuration the $d_z^2$ orbitals are alternately oriented along both [101] and $[10\bar{1}]$ directions, thus causing a Jahn-Teller distortion of the kind indicated in Fig. 9b.

*F-I Phase*

For the F-I ferromagnetic phase, Fig. 6 shows that the *a* and *b* axes increase slightly as the temperature decreases, while the *c* parameter remains practically constant. Resistivity measurements on the $x$=0.47 sample, which contains about 50% of F-I down to 10 K, show that this phase is a conductor also at low temperatures, as would be expected on the basis of a percolation model. The Mn-O distances at room temperature (Table 2) do not show the presence of a coherent Jahn-Teller effect. As the temperature decreases below 150 K, however, a distortion of the apical compression type appears and, at 10 K, with Jahn-Teller parameter[15] $\sigma_{JT}=1.4\times10^{-2}$. This behavior may be associated with a partial orbital ordering taking place in this phase, as shown in Fig. 9c and d.

The lattice parameters and ordered moments for both the F-I and A-II phases are plotted in Fig. 10 as a function of composition. At 300 K, the lattice parameters and unit cell volume of the F-I decrease as the value of $x$ increases from 0.47 to 0.53. This result is expected since the proportions of $Ca^{2+}$ and $Mn^{4+}$ (whose ionic radii are smaller than those of $La^{3+}$ and $Mn^{3+}$, respectively) increase as $x$ increases. The behavior of the lattice parameters at 10 K, however, is more complex, since *b* increases and *c* decreases with increasing $x$, while *a* and the unit cell volume $V$ show a maximum at $x$=0.5. We may speculate that these results are associated with the partial orbital ordering invoked previously to explain the Jahn-Teller effect observed in the F-I phase at low temperature.

***Magnetic field dependence***

The properties of CMR systems are extraordinarily sensitive to the application of magnetic fields, and the present samples are no exception, affecting both the crystallographic and magnetic phases. The discontinuous nature of the structural transformation between the F-I and A-II phases manifests itself in a dramatic difference in the state obtained depending on whether the sample is cooled in a field, or cooled in zero field and then the field applied. The refinement results obtained on the $La_{1/2}Ca_{1/2}MnO_3$ sample are given in Table 3 for a series of field and temperature paths. In no case did we detect any tendency for the particles themselves to rotate in the applied field, and produce a preferred orientation of the powder. Hence we were able to refine the nuclear structures for the F-I and A-II phases and obtain the phase fractions. Fig. 11 shows the temperature dependence of the lattice parameters obtained in a field of 5 T, where we see that the overall behavior is quite similar to the behavior in zero field (Fig. 7).

The application of a 8.5 T field after cooling in zero field (column 2 of Table 3) slightly decreases the proportion of A-II (from 76 to 68% of the sample). We also find that the refined value of the *z*-component of the $Mn^{3+}$ moment increases from 0.5 to 1.3 $\mu_B$, with a corresponding decrease of the *x*-component (from 2.3 to 1.6 $\mu_B$). However, it should be noted that in the geometry of the experiment the applied field is always perpendicular to the reciprocal lattice vector, hence at a given field there is a different and unknown distribution of angles that the field makes with the spins for each reflection. With the complications of two magnetic phases and a limited number of magnetic peaks, it is not possible to obtain reliable values of the moments. Hence these variations in the refined moments must simply be regarded as trends.

When a magnetic field of 8.5 T is applied during cooling from room temperature to 5 K, the formation of the A-II phase is strongly inhibited and the value of $\mu_x$ is dramatically reduced compared to the zero-field cooled state. In particular, after turning off the field, the *x*-component of the $Mn^{3+}$ moment is 1.0 $\mu_B$ (column 5 of Table 3), while for data collected in a field of 8.5 or 5.0 T $\mu_x$ completely disappears. The magnitude of $\mu(Mn^{3+})$ is only slightly higher than about 1.0 $\mu_B$ in all the experiments, in which the field is applied on cooling.

Although the magnetic field applied during cooling strongly favors the retention of F-I, which constitutes about 75% of the sample (instead of about 25% when no field is applied), this phase is clearly in a metastable state, and in fact heating the sample from 5 K to 25 K under no applied field (column 6 of Table 3) quickly re-establishes the zero-field-cooled state. On the other hand, the magnetic moment of the $Mn^{4+}$ cations is only slightly affected by the sample treatment, and that of F-I is generally higher than 3.0 $\mu_B$ when a field is applied during the data collection at 5 K.

The overall behavior of the system in an applied field does not exhibit any unusual or unexpected features, but it does emphasize that magnetic, structural, and electronic properties are in a delicate balance that can be dramatically shifted with only small external perturbations.

## Discussion

The overall behavior of $La_{½}Ca_{½}MnO_3$ at half-filling is summarized in the phase diagram of Fig. 1. At room temperature we have a single crystallographic phase, which orders ferromagnetically at 265 K. A second crystallographic phase begins to form below ~230 K, and the structures of these two phases are quite similar, but the Mn-O octahedron distorts differently in the two and the lattice parameters are distinctly different. The intrinsic widths of the Bragg peaks in this temperature regime argue that these distorted phases develop microscopically. Indeed, we attempted to physically separate the two phases by cooling the x=0.50 sample with dry ice to ~200 K, where the two phases are approximately in equal proportions (Fig. 4), and then removing the ferromagnetic portion with a permanent magnet. However, the *entire* sample was lifted by the magnet, and with the large and finely powdered sample it seems very unlikely that all the grains would be ferromagnetic if the two phases existed macroscopically, due to extrinsic defects or inhomogeneities. The data presented here indicate that the two phases do indeed coexist microscopically, and they transform discontinuously from one to another as a function of temperature or magnetic field.

The coexistence of two (or more) phases has been observed by a number of other groups. Xiao *et al.*[16] found as a function of applied field a series of distinct (discontinuous) phases with dramatically different conductivities, in a single-phase sample (at room temperature). Chen and Cheong[5] found in electron diffraction measurements the formation of an incommensurate phase that would correspond to our 230K structural transition. This new phase exhibited hysteresis and two-phase coexistence, and these conclusions were consistent with subsequent NMR, infrared, and transport data.[17-19] Our results are consistent with these studies, as well as with recent diffraction data. However, we find that the incommensurate phase develops into a distinct second phase rather than remaining as a charge density wave with discommensurations. The F-I and A-II phases then coexist, with regions transforming between the two phases discontinuously as a function of temperature and applied magnetic field. The ferromagnetic phase itself, however, does not exhibit any tendency to cant, but instead maintains its collinear arrangement with the easy axis along *c*.

The nuclear and magnetic structural results obtained in our analysis are generally in good agreement with published data[6,9] and with Goodenough's theoretical predictions[9]. With our experimental conditions, however, we have not been able to observe in the neutron diffraction patterns taken below $T_N \cong 160$ K the satellite peaks generated by the modulation discussed by Radaelli et al.[8], or to detect the monoclinic distortion required by the charge ordered state associated with the CE magnetic structure.

The behavior of the Mn-O distances as function of temperature, illustrated in Fig. 8, shows that in both the F-I and A-II phases orbital ordering occurs, and the transition from the disordered state to the ordered Jahn-Teller state takes place in stages. In the range between 220 and 150 K, *i.e.* in the same interval in which the diffraction peaks are significantly broadened, the Mn-O distances change rapidly with temperature. The overall distortion of the $MnO_6$ octahedra is different in the F-I and A-II phases, and they coexist to low temperatures. This result is completely consistent with Radaelli's interpretation of the peak broadening in terms of co-existing phases having similar lattice parameters.

Finally, we note that Radaelli et al.[8] reported evidence in their x-ray and neutron diffraction patterns of the presence of the F-I phase down to 1.5 K, where it accounts for about 7% of the total integrated intensity. These authors considered a small magnetic peak observed at 1.5 K to be typical of the G-type antiferromagnetic structure, discussed by Wollan and Koehler[6], and attributed it to the F-I phase. However, a comparison of Fig. 2b and 2c shows unambiguously that the F-I at 10 K retains the same ferromagnetic ordering observed at 175 K, and there is no G-type magnetic order in the system.

## Acknowledgments

We would like to thank S-W. Cheong, D. E. Cox, and P. G. Radaelli for helpful discussions. Research at the University of Maryland supported in part by the NSF-MRSEC DMR 96-32521.

## Figure captions

Fig. 1. Phase diagram of $La_{1-x}Ca_xMnO_3$ in the range of compositions $0.47 \le x \le 0.53$. The horizontal curves separate, going from top to bottom: (i) the ferromagnetic (FM) transition of the F-I crystallographic phase at ~260 K; (ii) the formation of the low-temperature A-II phase, which appears at ~230 K; and (iii) the antiferromagnetic (AFM) transition, which occurs in the A-II structure at ~160 K ($T_N$). As shown in the diagram, the ferromagnetically ordered F-I phase and the antiferromagnetically ordered A-II phase coexist at low temperatures.

Fig. 2. Neutron powder diffraction pattern of the $x$=0.47 composition in the angular range between 15 and 40° at various temperatures, employing the Cu(311) monochromator. In (a) only the nuclear peaks of the F-I phase are present. (b) the ferromagnetic reflections associated with the F-I phase are indexed in the difference plot, obtained by subtracting from the observed intensities the calculated structural profile. Note that at 175 K only the magnetic and nuclear peaks of F-I clearly visible, since the fraction of A-II phase is small and the peaks are broad at this temperature. (c) at 10 K both the ferromagnetic peaks of the F-I phase and the antiferromagnetic peaks of the A-II phase contribute to the diffraction pattern, and these peaks are clearly visible in the difference plot. Note that the ferromagnetic peaks at 10 K appear at the same angular positions as those observed at 175 K. (d) complete fit of the neutron pattern taking into account all nuclear and magnetic intensities. The small peak at $2\theta \cong 20°$ is the second order of the strongest nuclear reflection.

Fig. 3. Observed (crosses) and calculated (solid curve) neutron profiles for the $x$=0.50 composition at 10 K, with the data taken with the Ge(311) monochromator. Both the F-I phase with ferromagnetic order and the A-II phase with antiferromagnetic (CE-type) order are necessary to adequately account for the observed intensities in the refinement. The short vertical lines mark the $2\theta$ angles of the Bragg reflections for F-I (middle), A-II (bottom), and the CE-type ordering (top). The difference plot, at the bottom of figure, is obtained by subtracting from the observed intensities the calculated profile. The inset shows a plot of the (031), (211) and (112) reflections at a series of temperatures. Below ~230 K the (031) of the A-II phase begins to separate, and sharpens and shifts towards higher angles as the temperature decreases.

Fig. 4. (a) Full width at half maximum (FWHM) of the $(031)_{A-II}$ nuclear reflection as function of the temperature. (b) integrated intensity of $(031)_{A-II}$ reflection showing that the A-II contribution to the diffraction intensities increases as the Ca content increases. (c) phase fraction of A-II as a function of the temperature and composition. The curve for $x$=0.50 at 5 T shows that the application of a magnetic field inhibits the formation of A-II (see text). In this figure the full symbols indicate results obtained on cooling and the open symbols results obtained on warming, while the solid curves are a guide.

Fig. 5. Temperature dependence of the resistivity and bulk magnetization on the identical samples used to collect the neutron data. The resistivity is low for x=0.47 where the sample consists mostly of the ferromagnetic phase, and increases dramatically with x=0.50 and 0.53 where the fraction of the sample in the ferromagnetic phase is much less as indicated from the magnetization. This demonstrates that the ferromagnetic phase is metallic while the antiferromagnetic phase is insulating, and the mixture can be understood on the basis of percolative conductivity.

Fig. 6. Zero-field magnetic moments in the F-I and A-II structures as a function of the temperature for *x*=0.47 (a), 0.50 (b), and 0.53 (c). The $Mn^{3+}$ and $Mn^{4+}$ moments in the A-II phase are roughly the same for *x*=0.47 and 0.50, while for *x*=0.53, the moment associated with $Mn^{3+}$ is lower than that of Mn4+ at all temperature below 150 K. In this figure the full symbols indicate results obtained on cooling and the open symbols results obtained on warming, while the solid curves are a guide.

Fig. 7. Temperature dependence of the lattice parameters for the F-I and A-II structures of the *x*=0.50 composition. The increase of *a* and *c* and the decrease of *b* with decreasing temperature indicate orbital ordering in *a-c* plane of the A-II phase (see text). The full symbols indicate results obtained on cooling and the open symbols results obtained on warming, while the solid curves are a guide.

Fig. 8. Variation of Mn-O distances as a function of the temperature in the F-I and A-II phases. The presence of the Jahn-Teller distortion below ~200 K in both structures indicates that orbital ordering begins to occur at the same temperature in both phases. In the temperature range between 200 and 150 K the behavior of the Mn-O distances indicates that a re-arrangement of the $\mathbf{d}_{\mathbf{z}}^2$ orbital takes place in both the F-I and A-II structures. The full symbols indicate results obtained on cooling and the open symbols results obtained on warming, while the solid curves are a guide.

Fig. 9. $MnO_6$ octahedra at 170 K and 10 K for the F-I (x=0.47) and the A-II (x=0.50) structures. The different Jahn-Teller distortions at 170 K and 10 K have been interpreted as due to a re-arrangement of the $\mathbf{d}_{\mathbf{z}}^2$ orbitals in both structures (see text and Fig. 8).

Fig. 10. Zero-field variation of lattice parameters, unit cell volume and magnetic moments of the F-I and A-II phases at 300 K and 10 K, as a function of the composition. In this figure the full symbols and the open symbols are just used for clarity of presentation.

Fig. 11. Temperature dependence of the lattice parameters of the F-I and A-II phases of the x=0.50 composition, measured on cooling (solid symbols) and warming (open symbols), in a magnetic field of H=5 T.

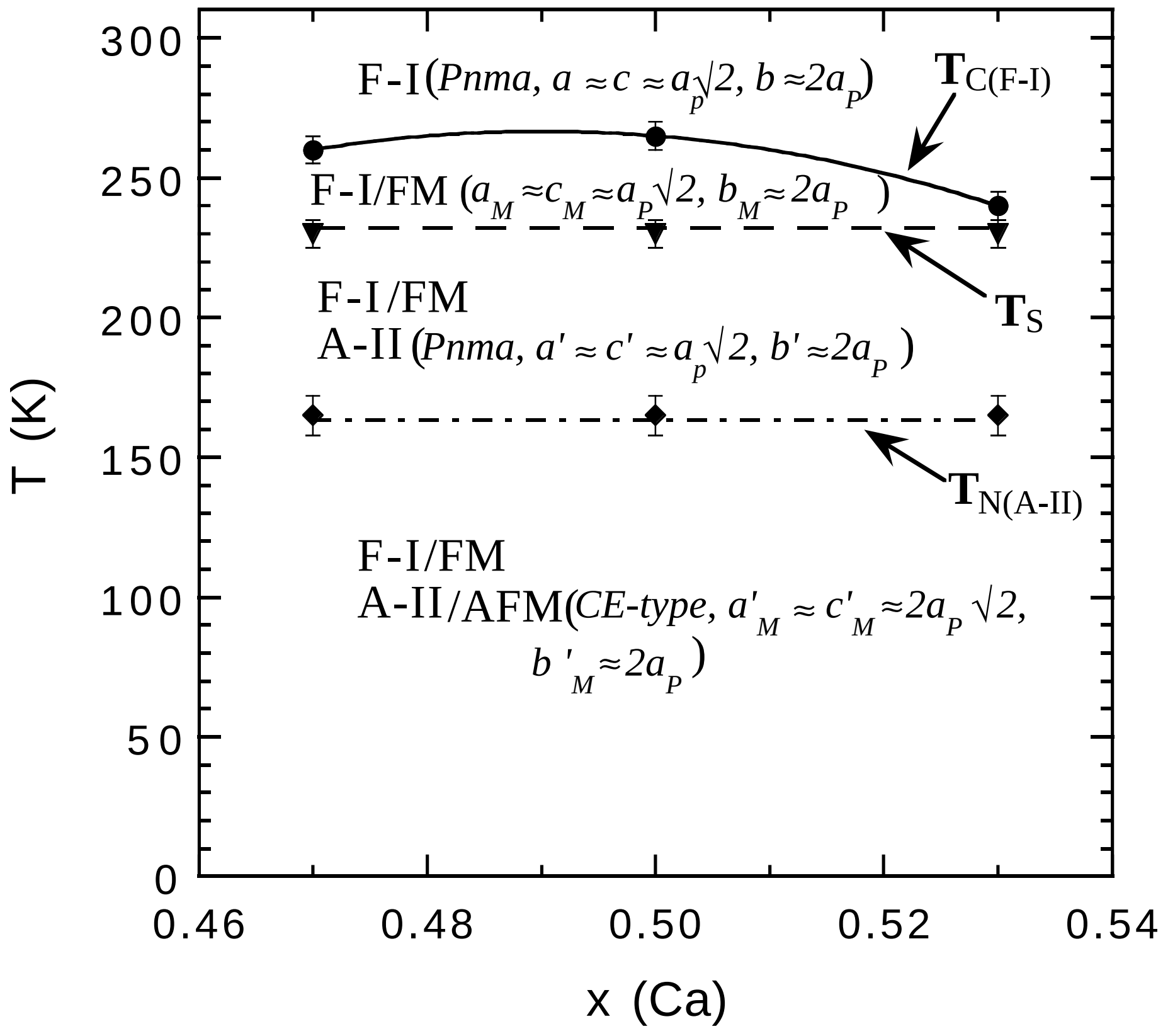

F-I(Pnma, a ≈ c ≈ a_p√2, b ≈ 2a_P)
T_C(F-I)
F-I/FM (a_M ≈ c_M ≈ a_P√2, b_M ≈ 2a_P )
T_S
F-I/FM
A-II (Pnma, a' ≈ c' ≈ a_p√2, b' ≈ 2a_P )
T_N(A-II)
F-I/FM
A-II/AFM(CE-type, a'_M ≈ c'_M ≈ 2a_P √2,
b '_M ≈ 2a_P )
T (K)
x (Ca)
300
250
200
150
100
50
0
0.46
0.48
0.50
0.52
0.54

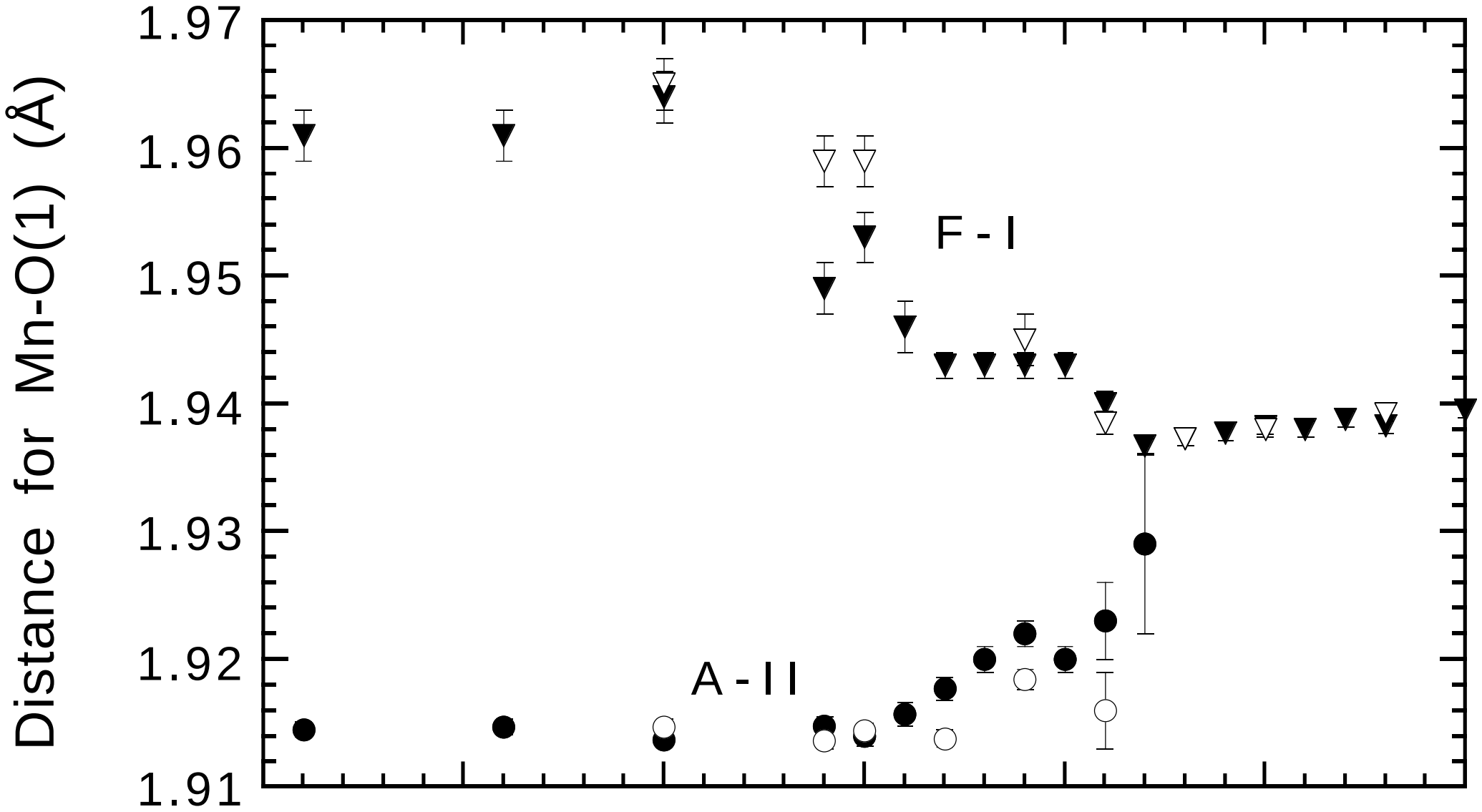

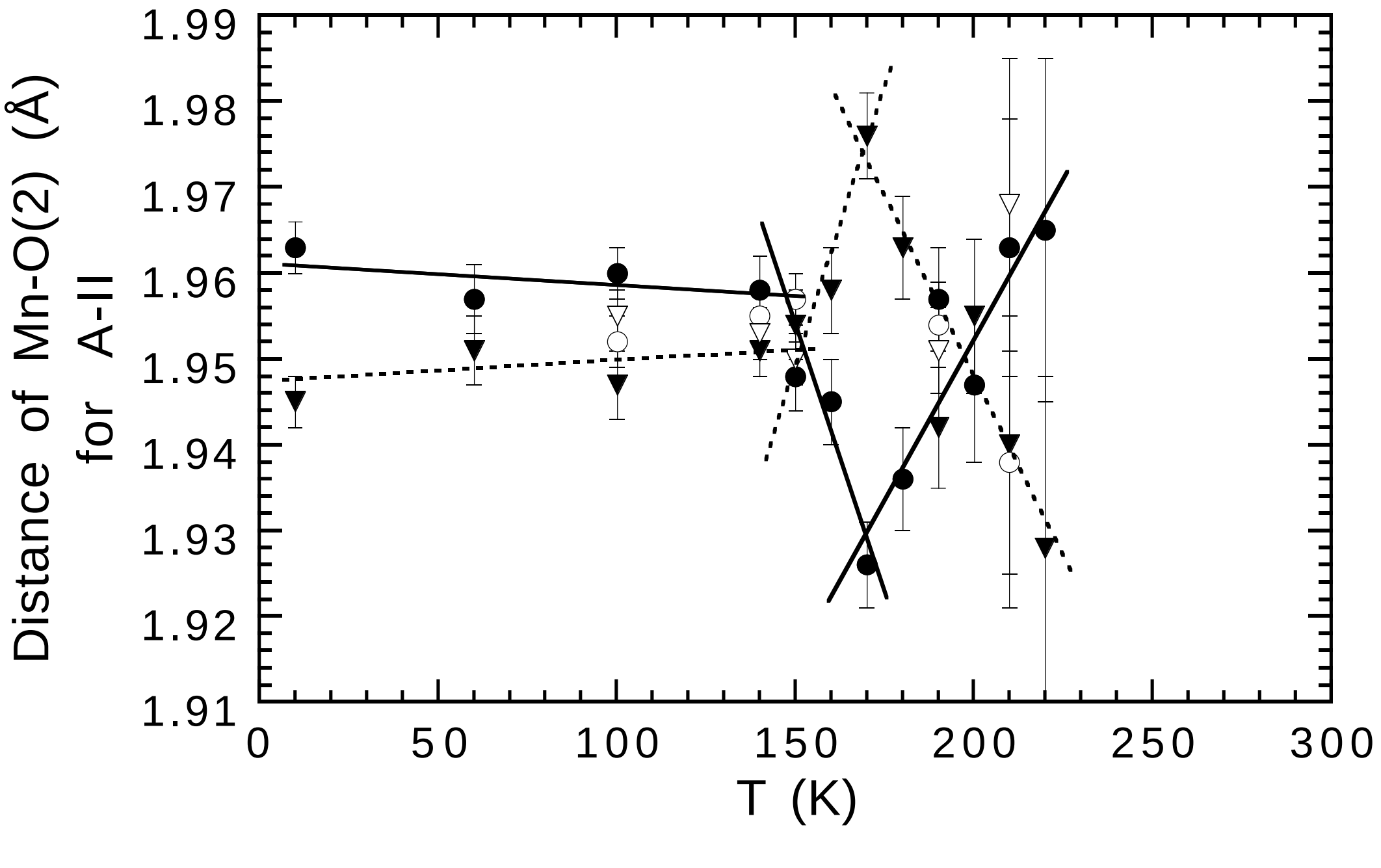

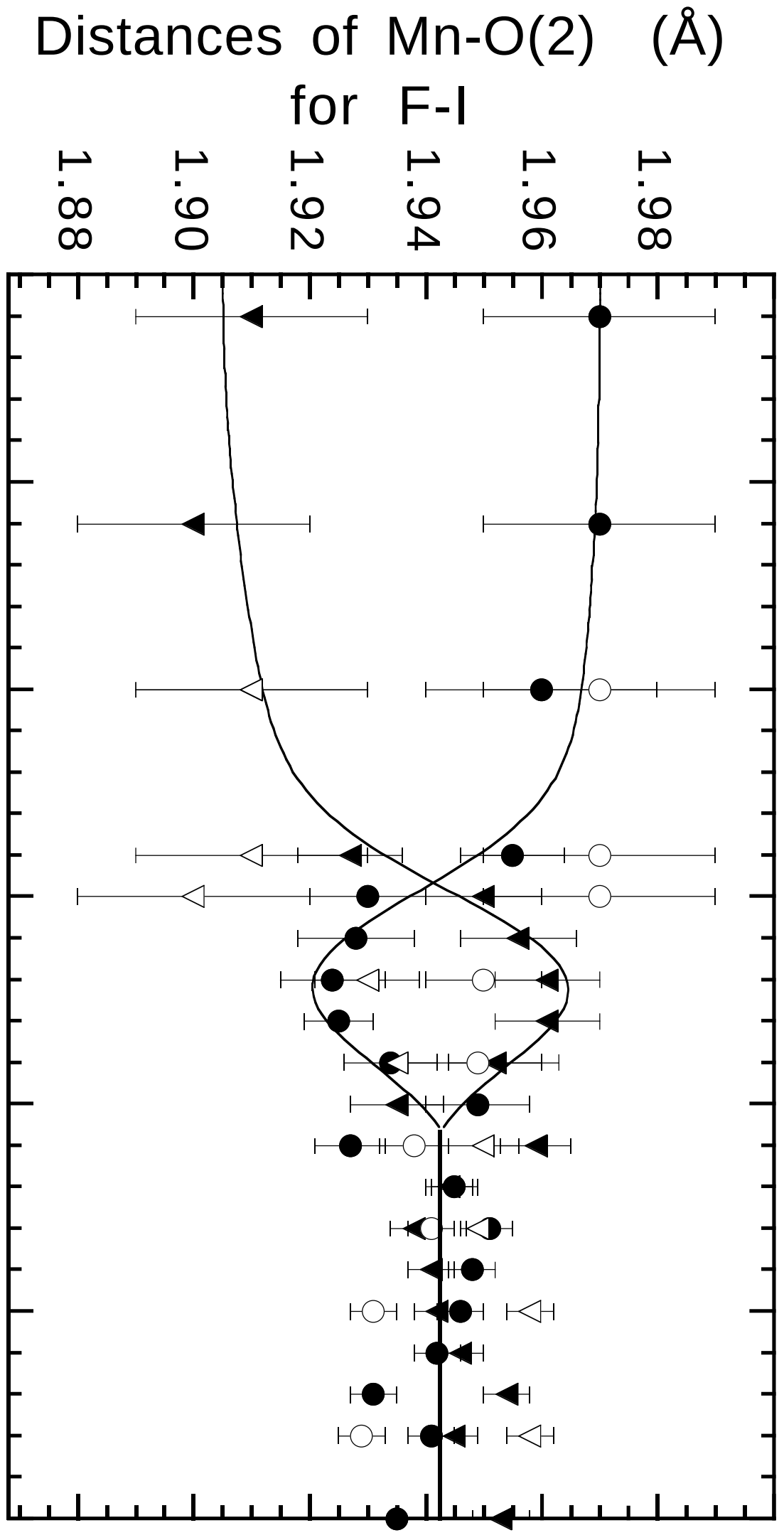

This figure "fig2.gif" is available in "gif" format from:

http://arxiv.org/ps/cond-mat/9912167v1

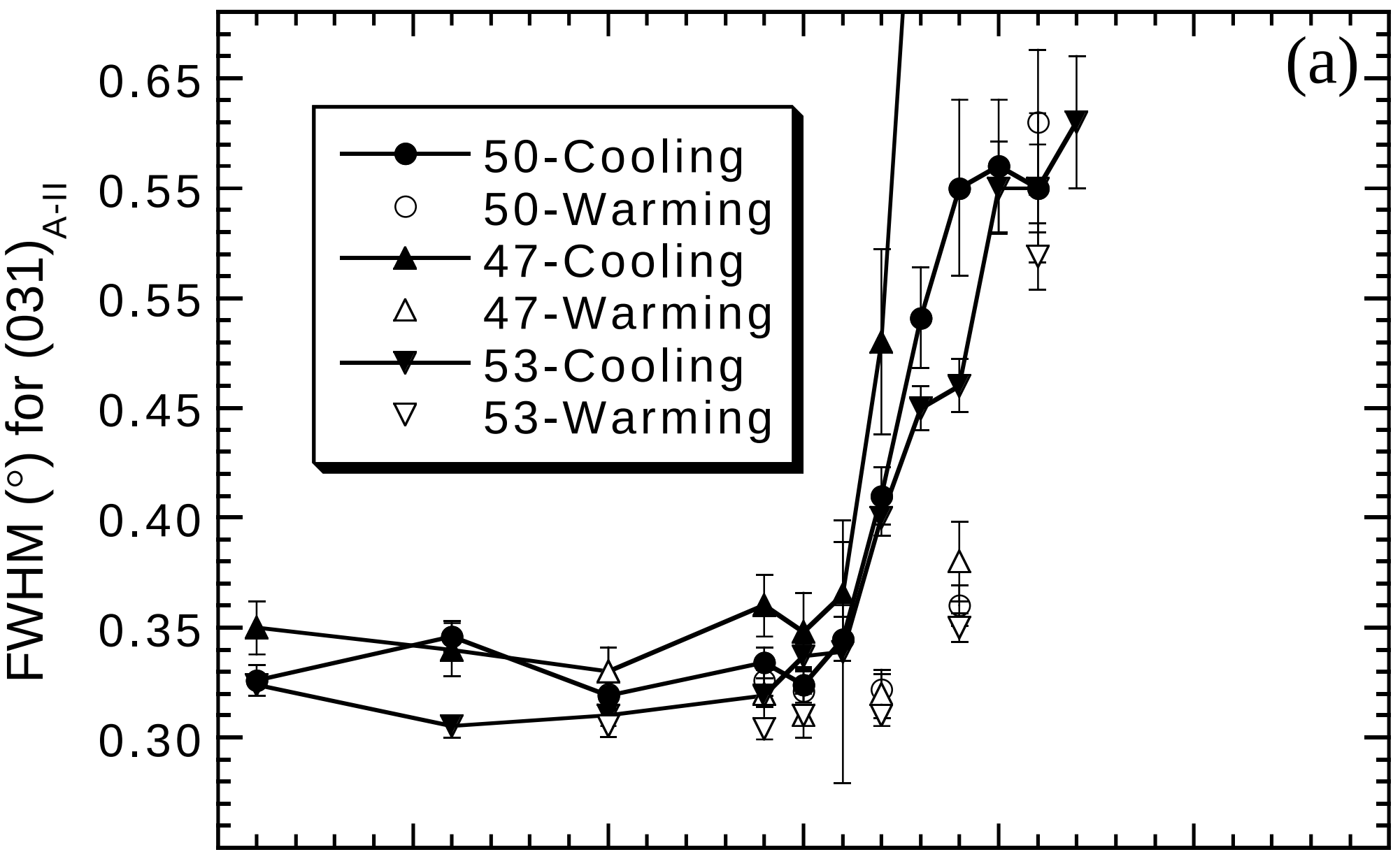

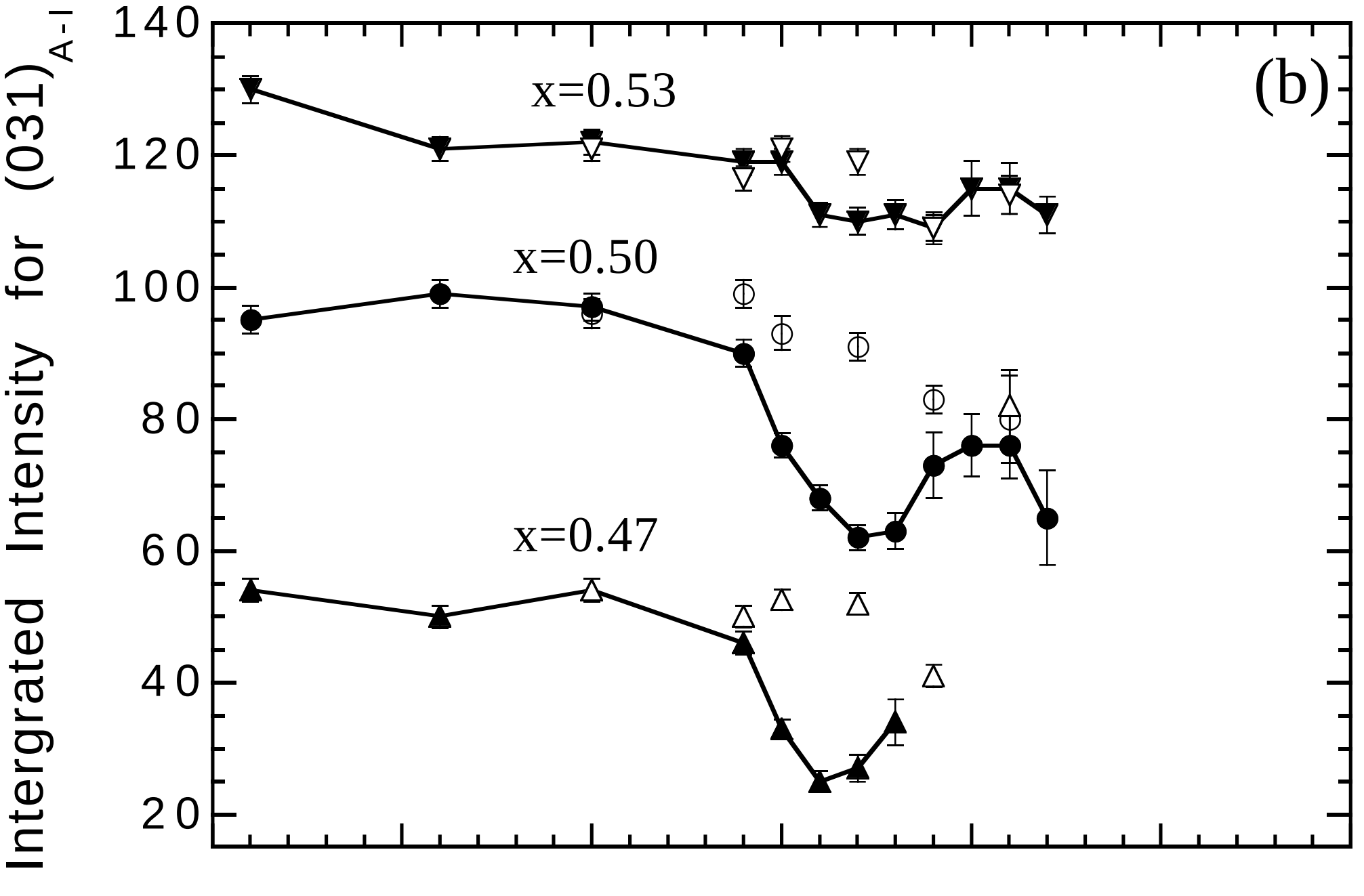

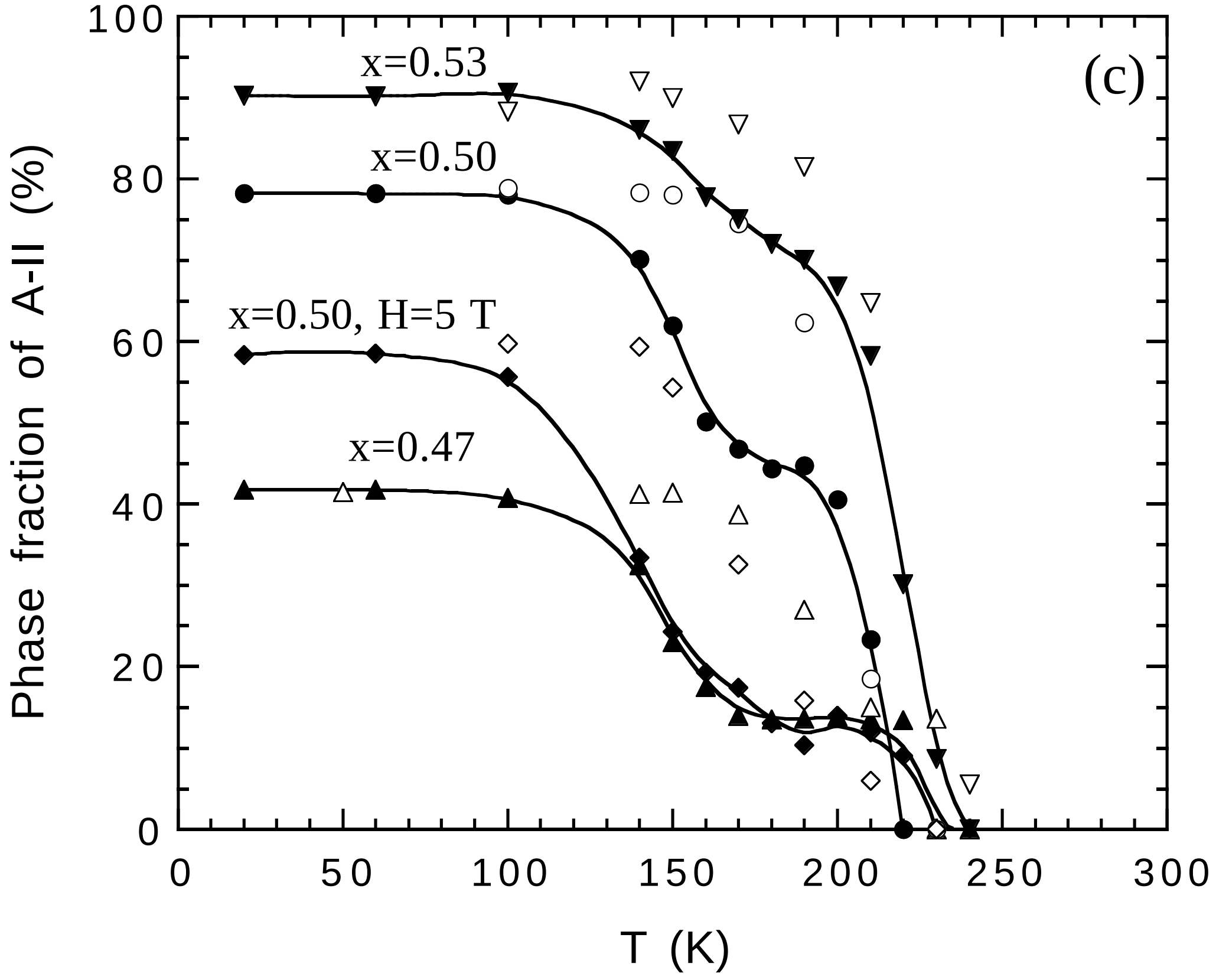

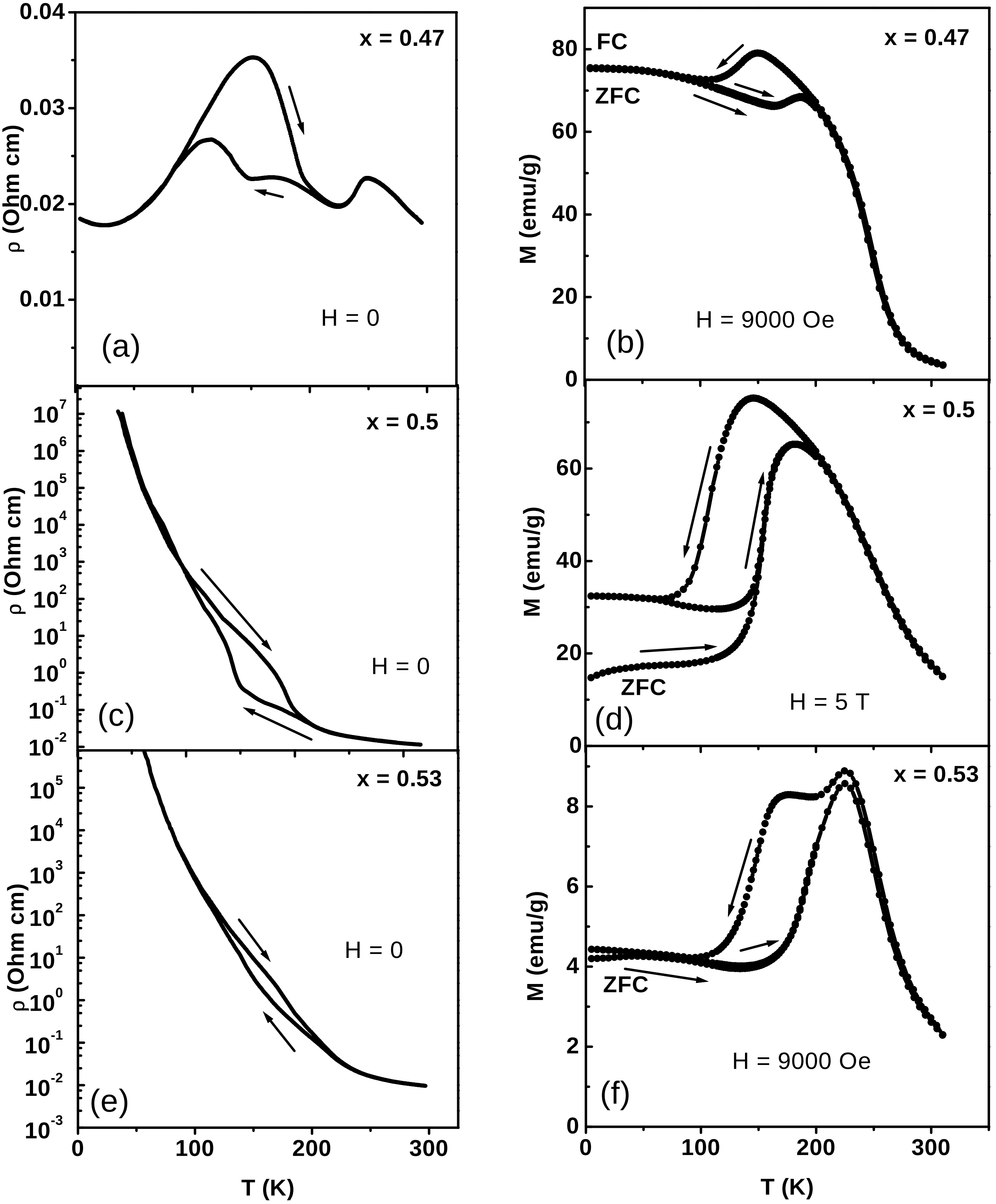

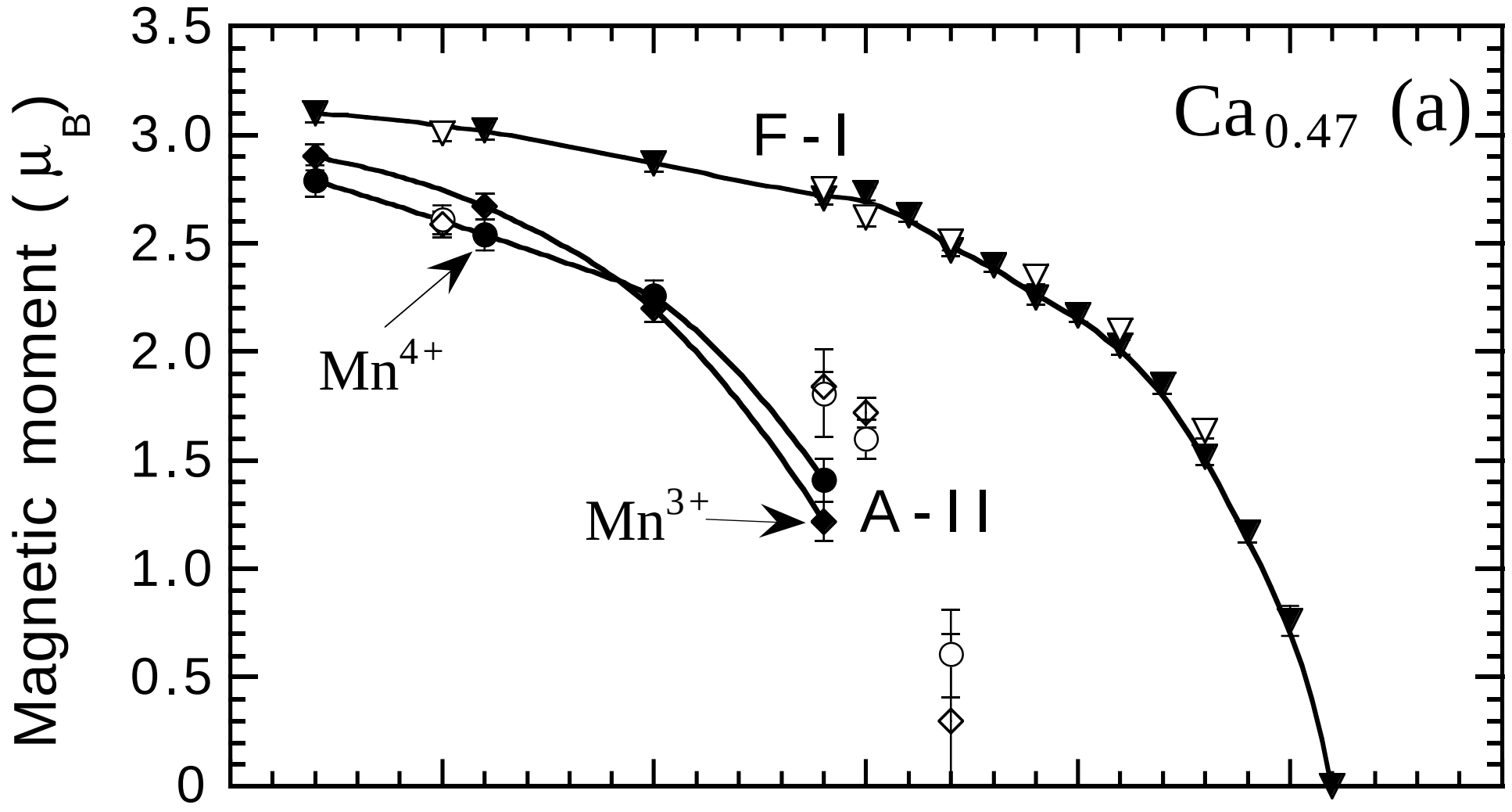

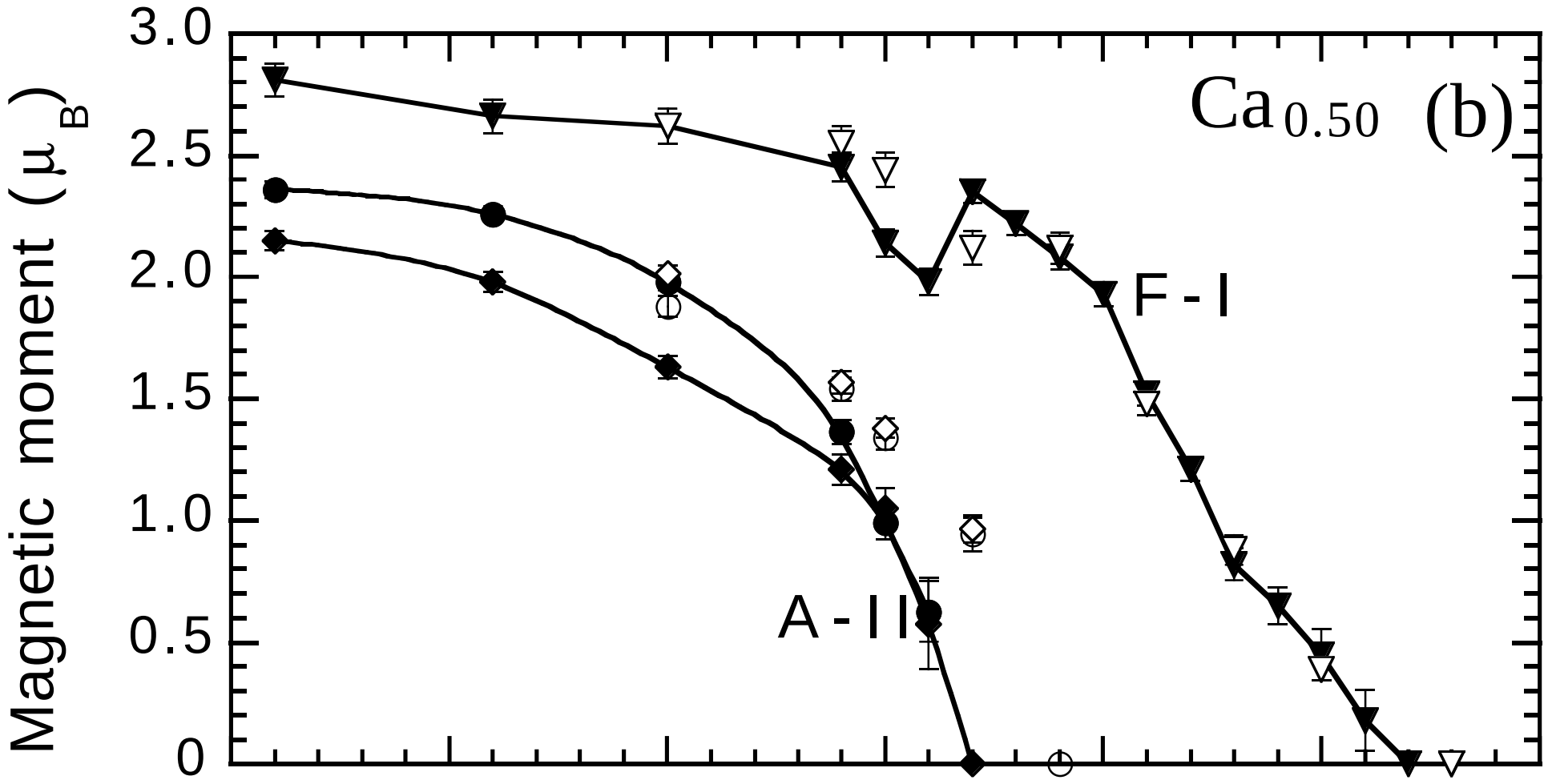

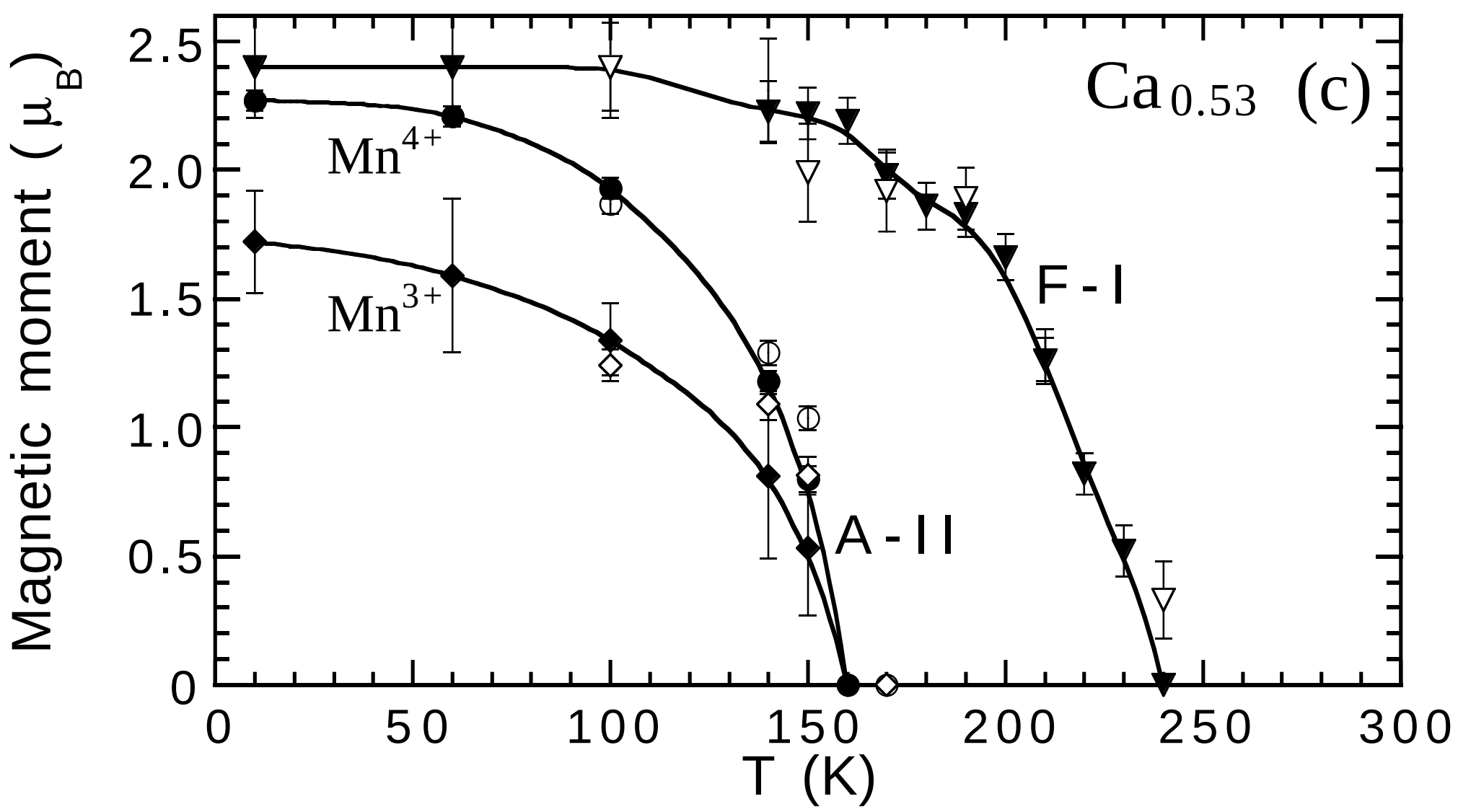

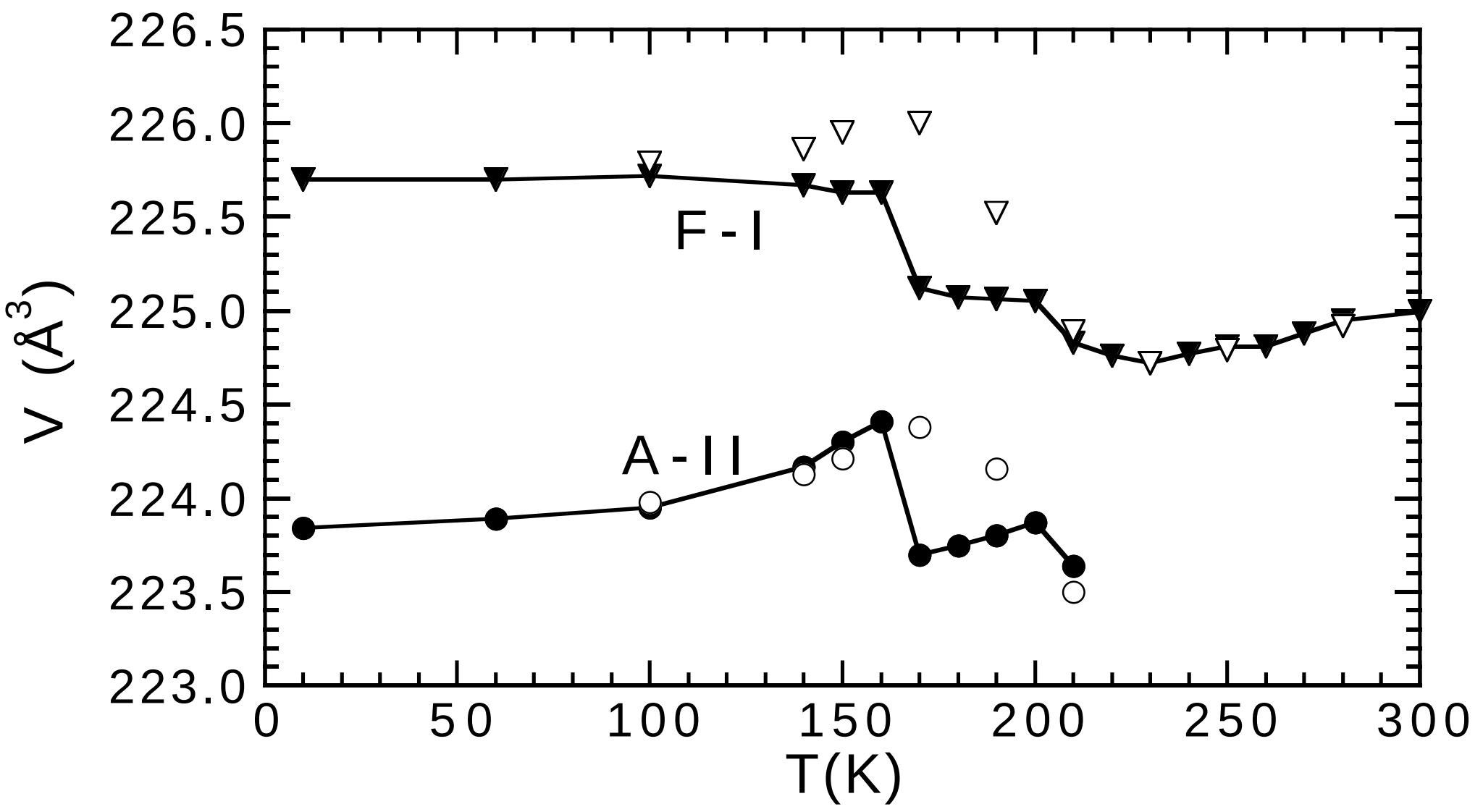

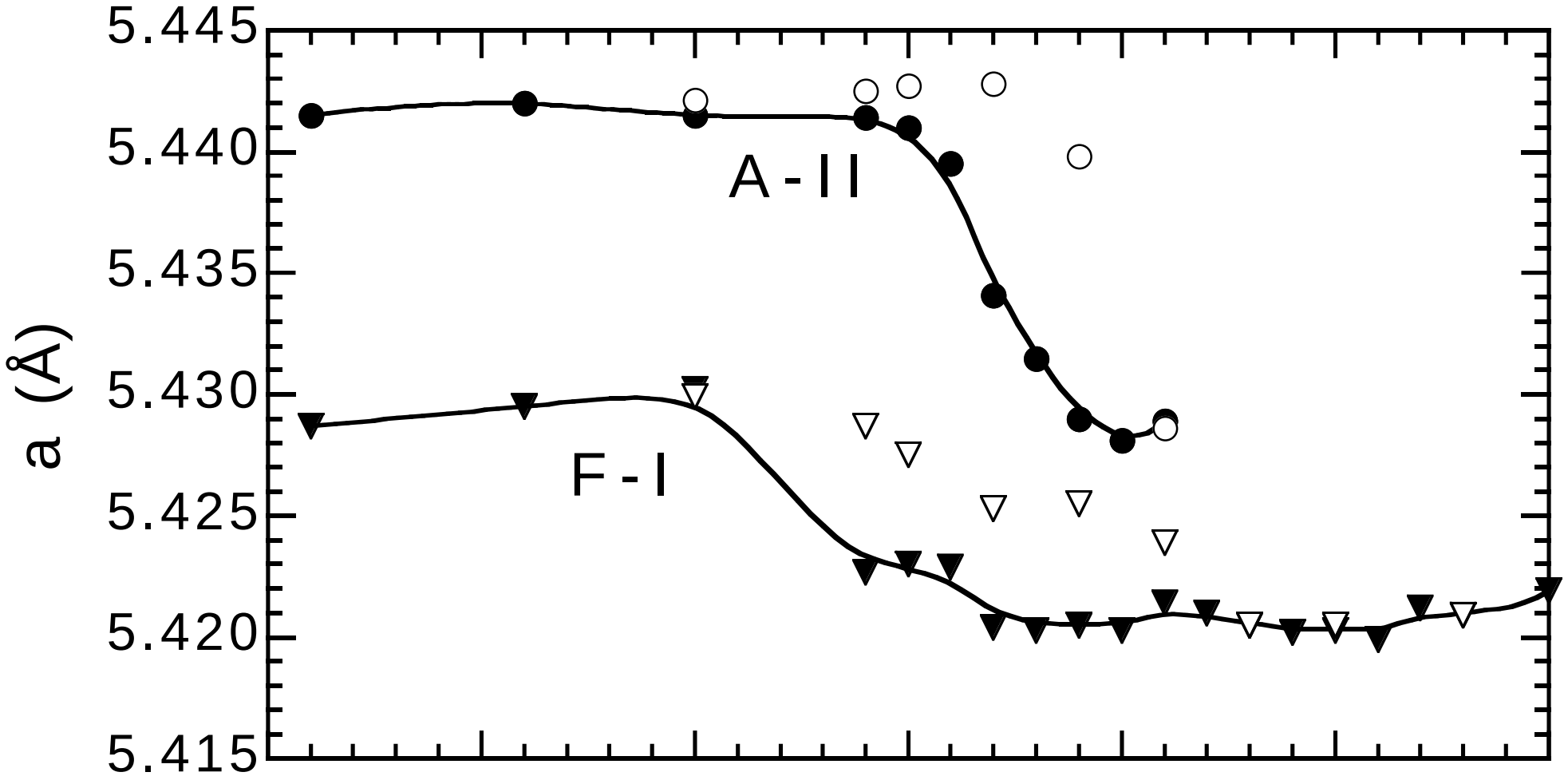

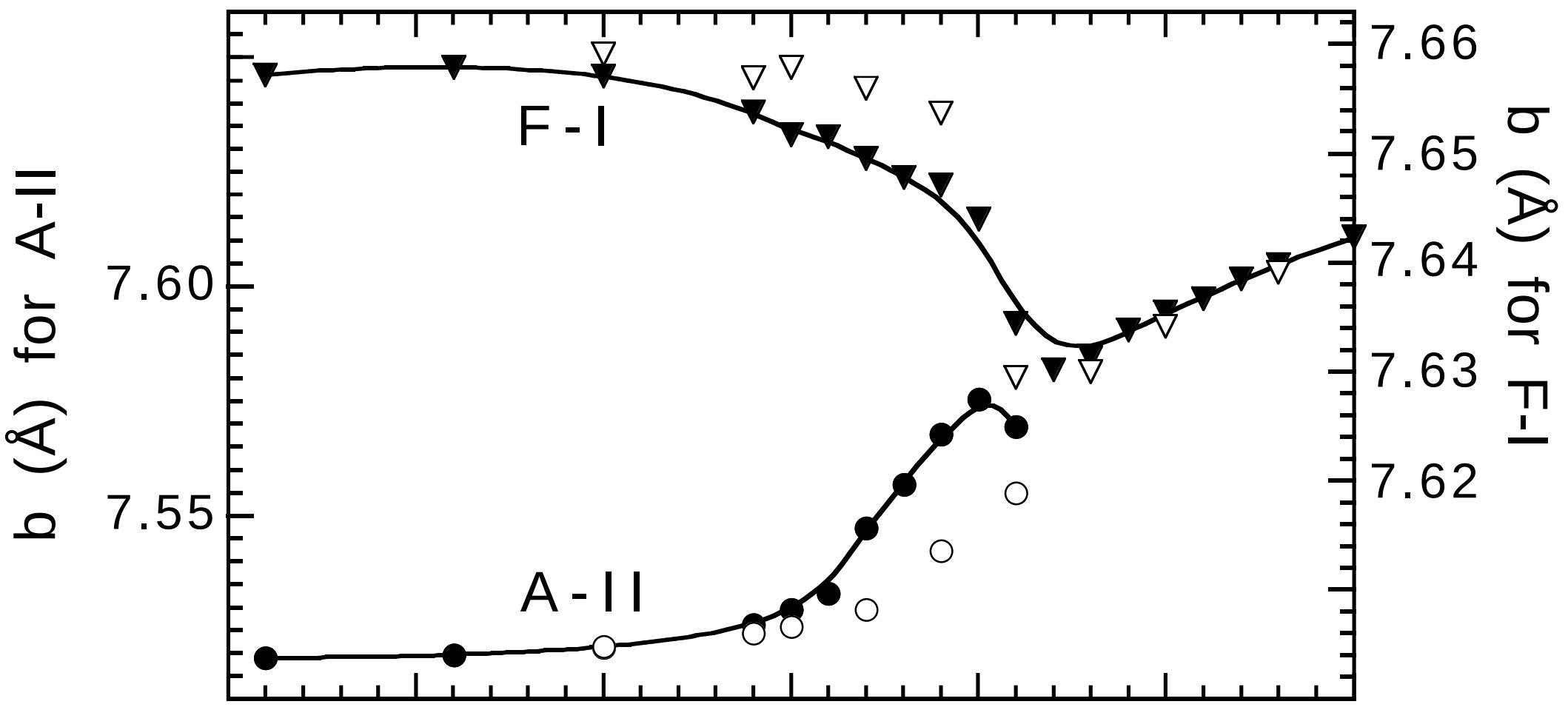

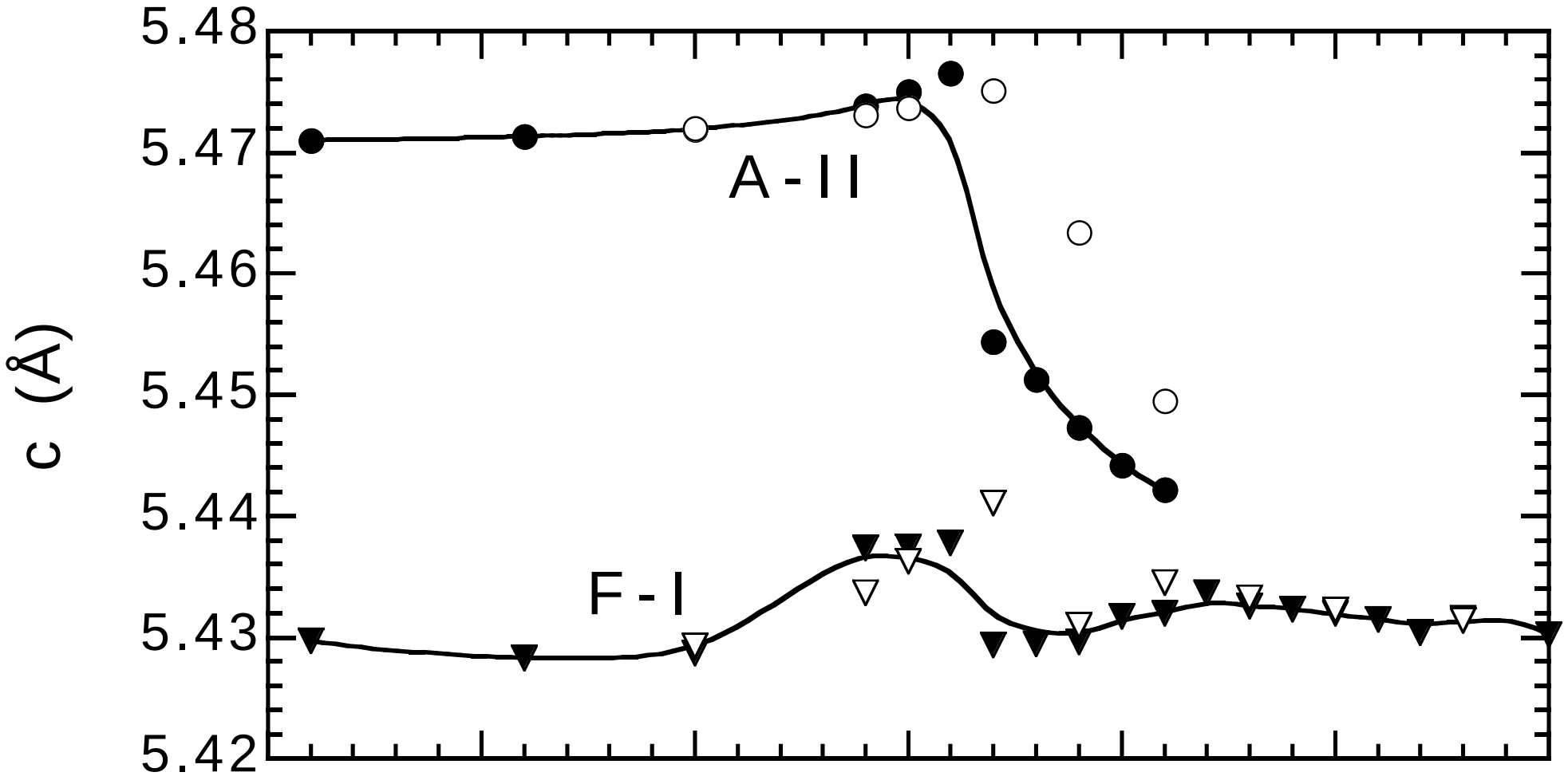

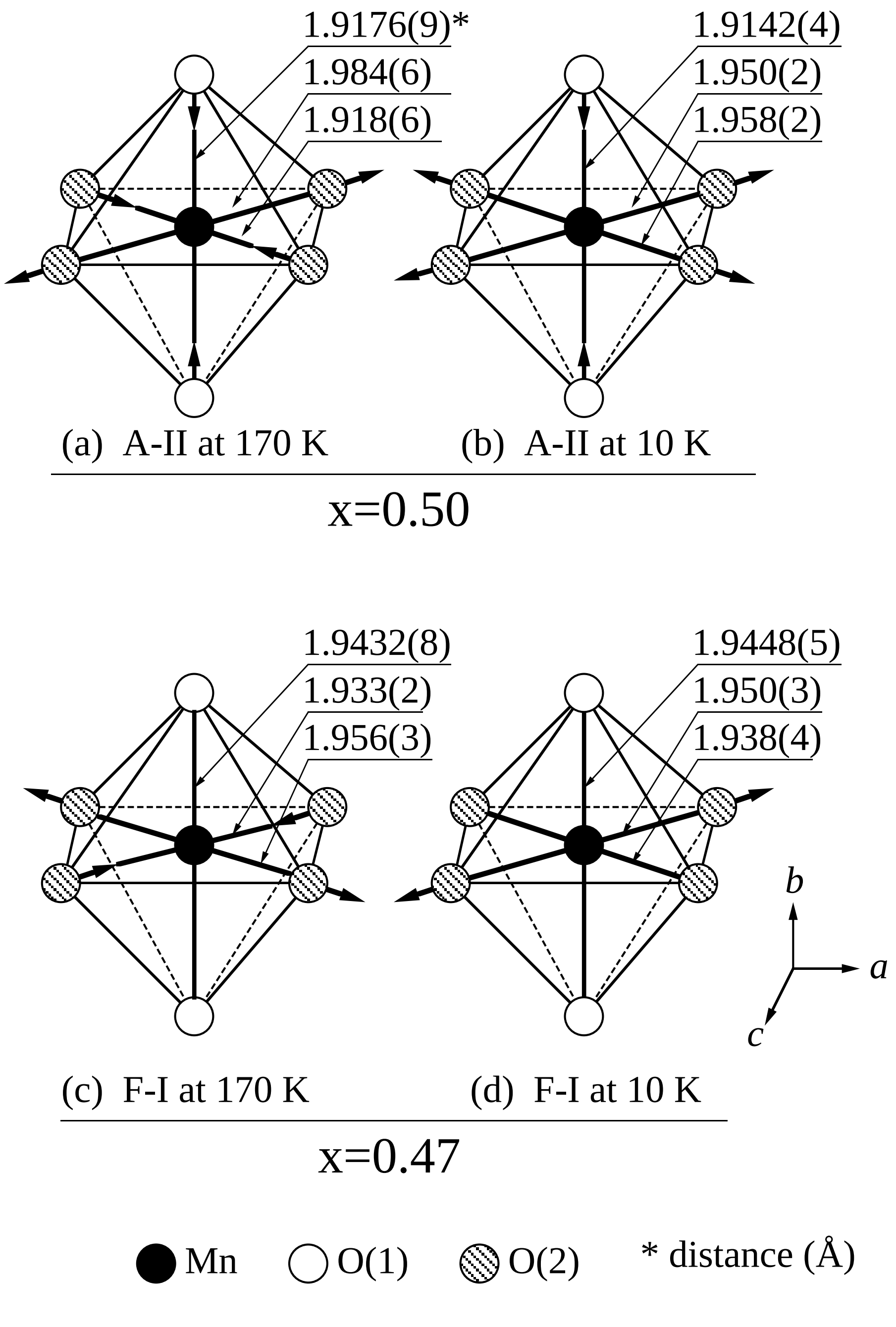

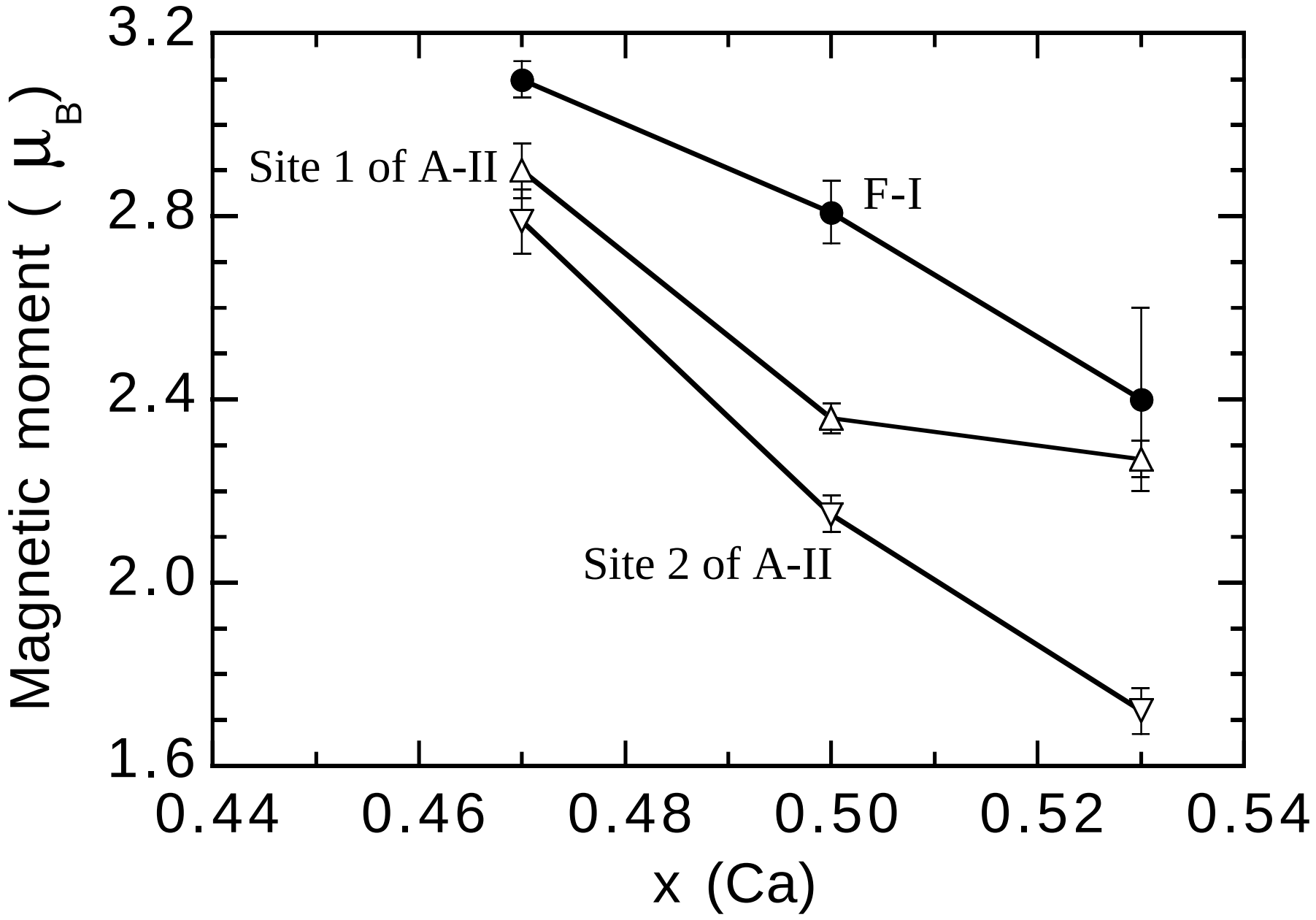

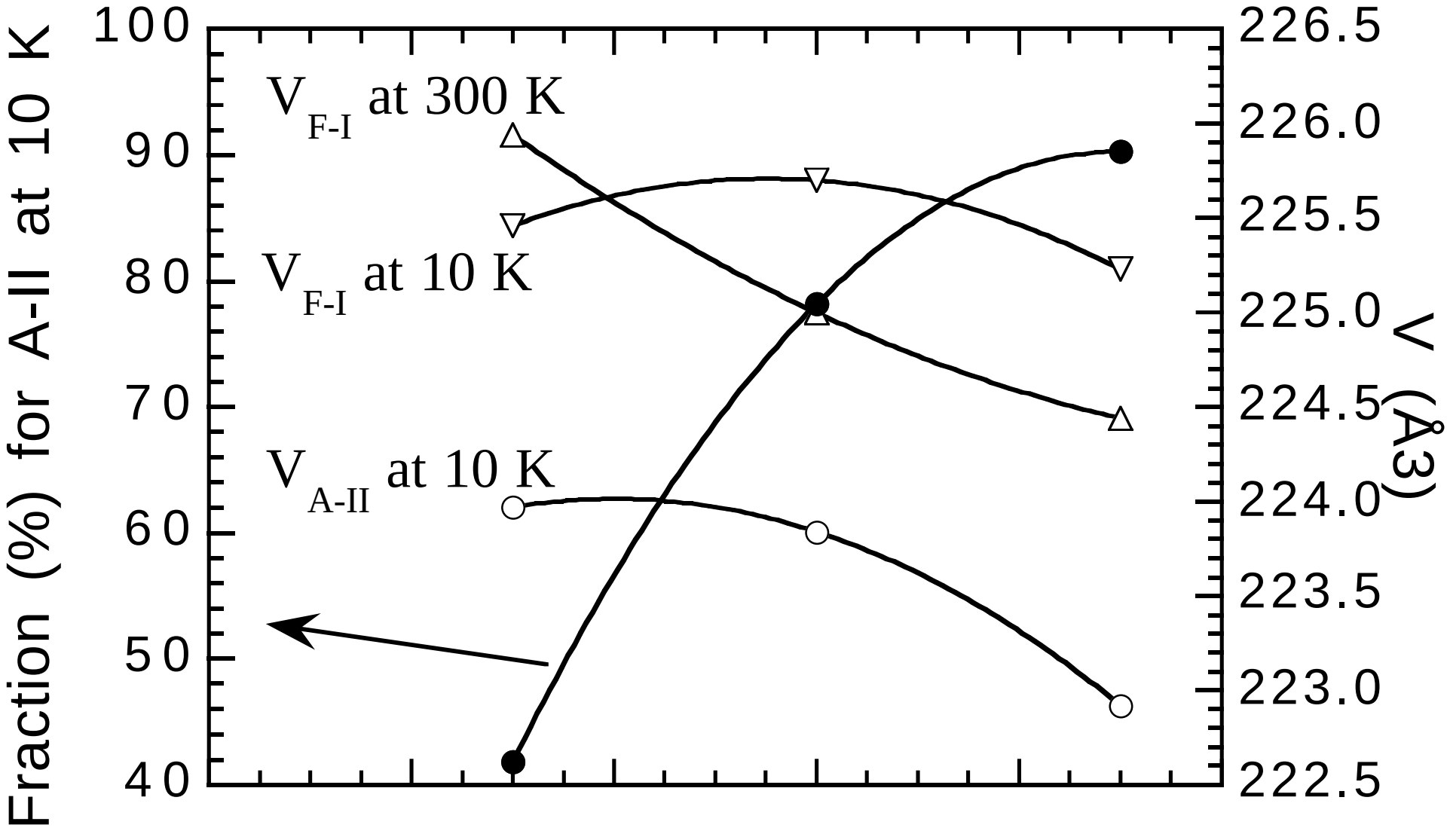

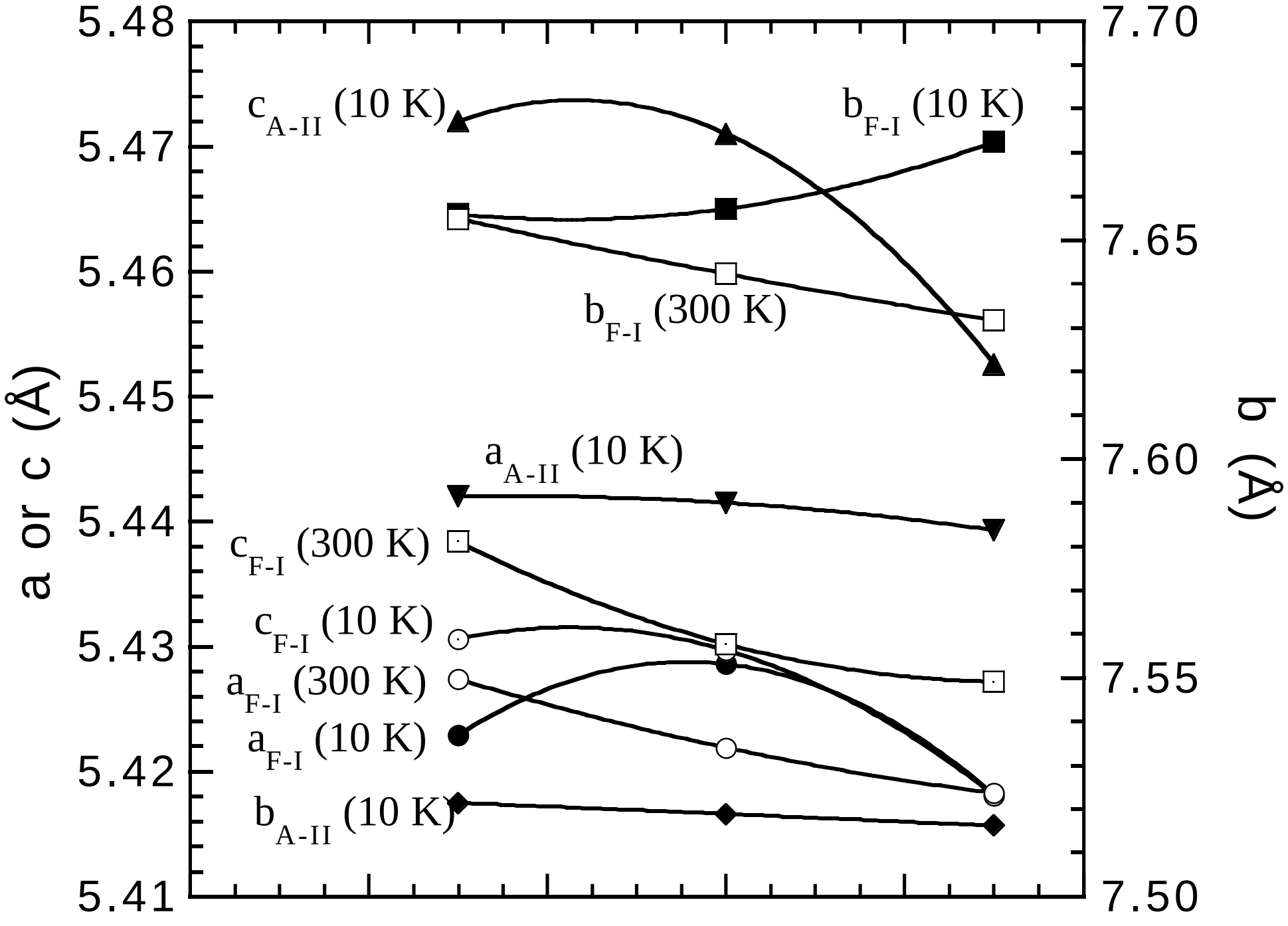

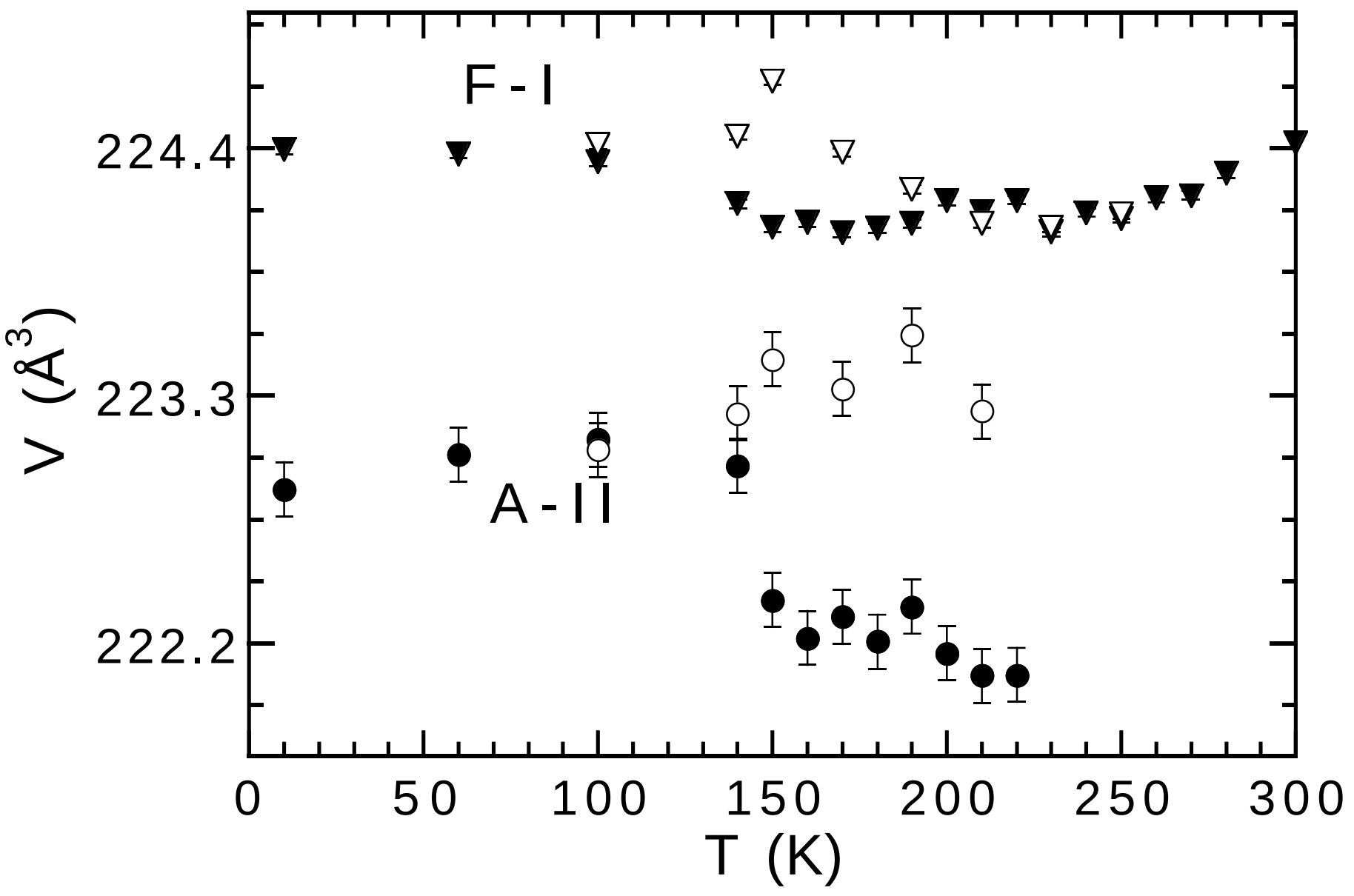

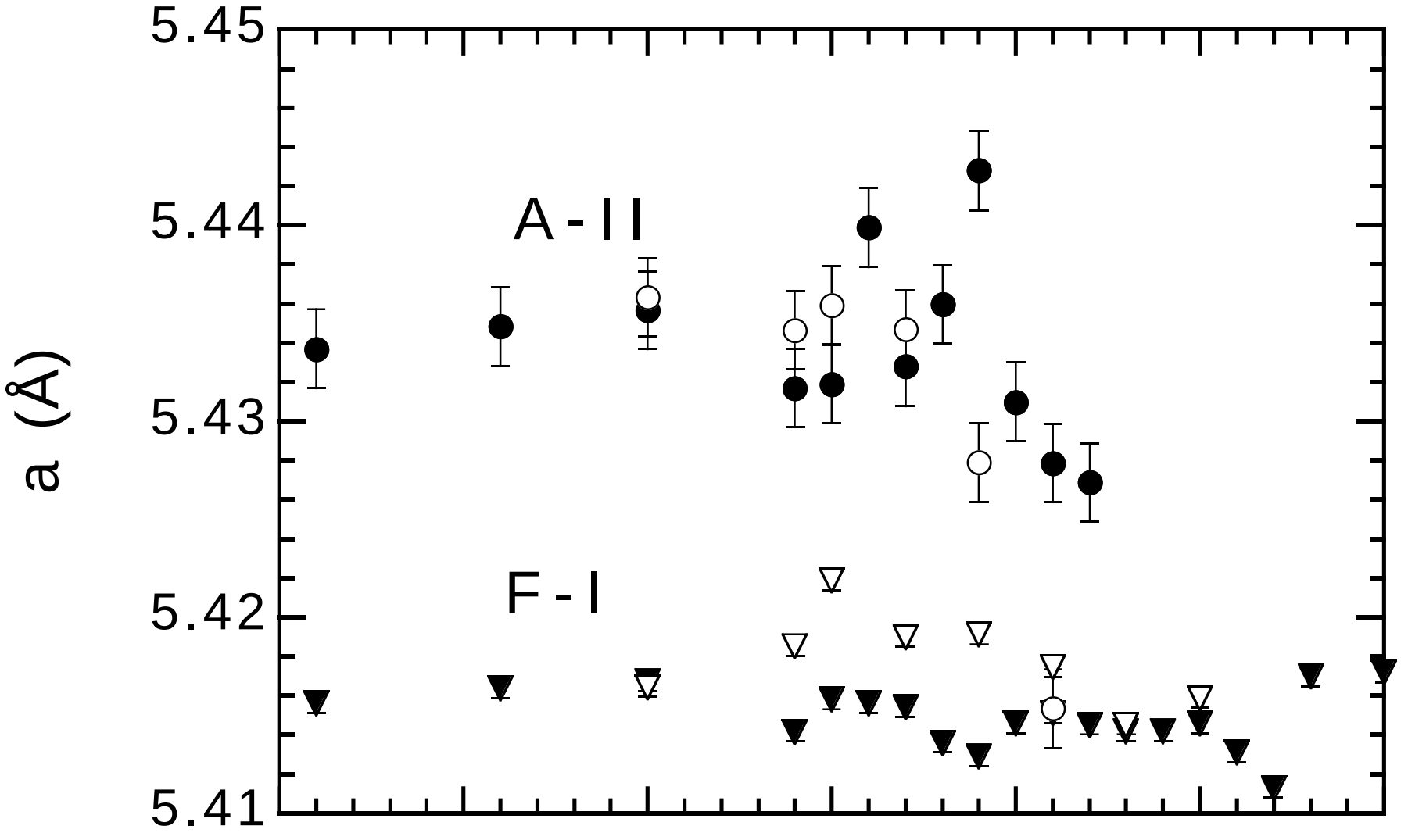

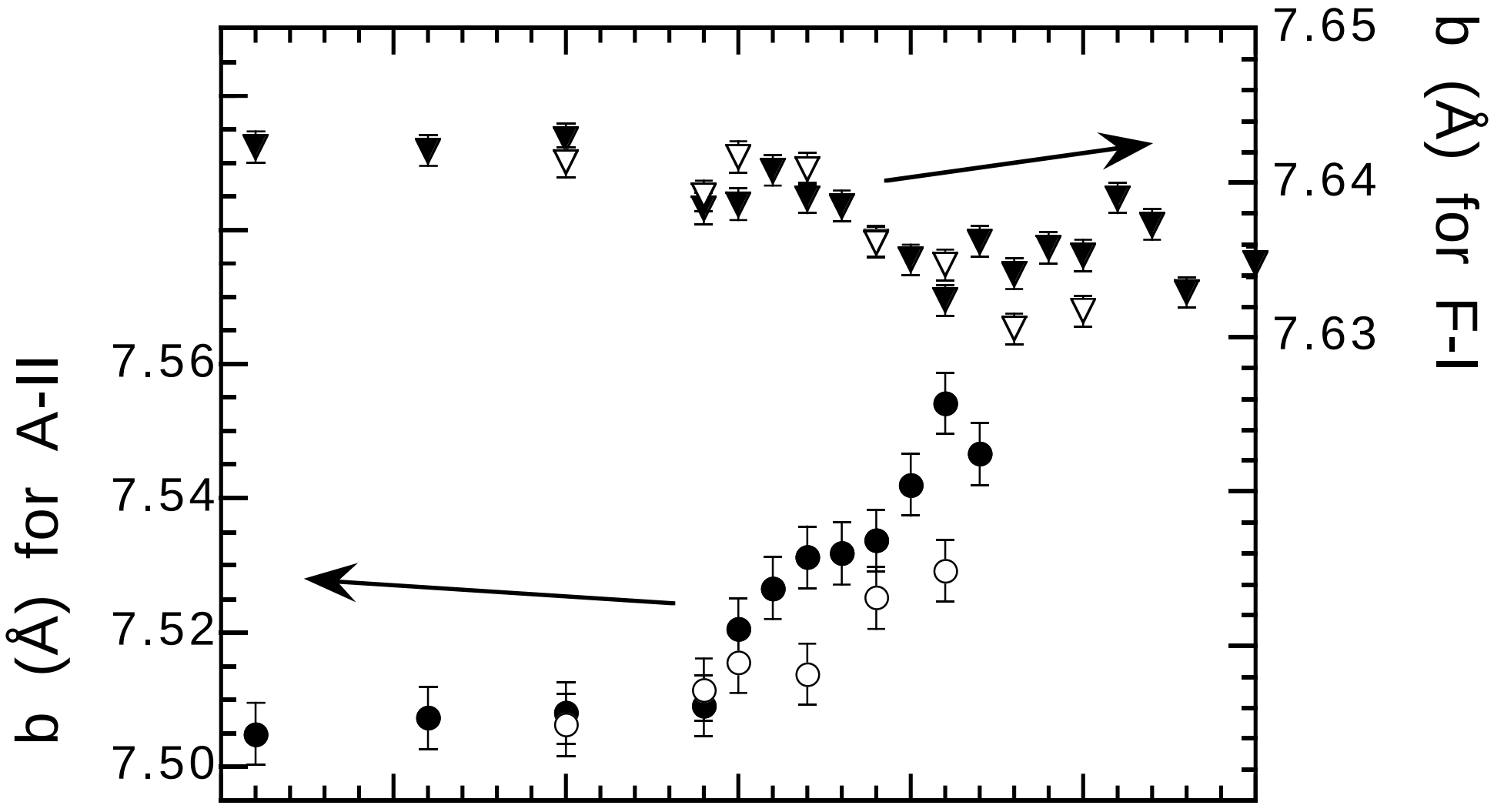

b (Å) for A-II
7.50
7.52
7.54
7.56
b (Å) for F-I
7.63
7.64
7.65

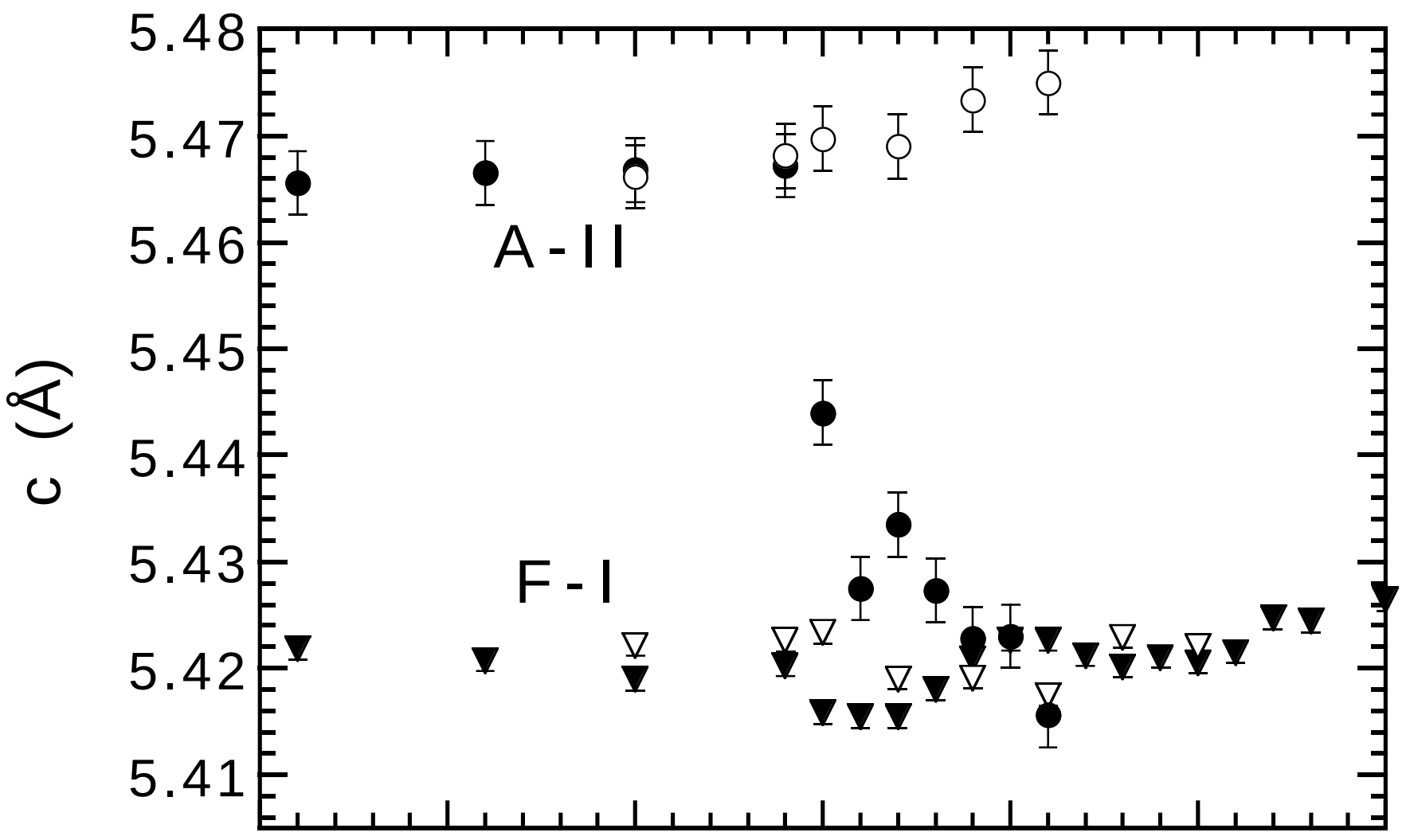